\documentclass[10pt, conference]{IEEEtran}

\usepackage[usenames, dvipsnames]{color}
\usepackage{url}
\usepackage{float}

\usepackage{breakurl}
\usepackage[hidelinks,breaklinks]{hyperref}
\usepackage{array}
\usepackage{color}
\usepackage{stmaryrd}
\usepackage[caption=false,font=footnotesize]{subfig}
\usepackage{graphicx}
\usepackage{lipsum,graphicx,multicol}
\usepackage{epsfig}
\usepackage{mdwmath}
\usepackage{mdwtab}
 \usepackage[super]{nth}
 \usepackage{nth}
\usepackage{bbold}
\usepackage{fixltx2e}
\usepackage{epstopdf}
\usepackage[table,xcdraw]{xcolor}
\usepackage{multicol, blindtext}
\usepackage{amsmath,bm}
 \usepackage{booktabs}
\usepackage{blindtext}
\usepackage{lipsum}
\usepackage{afterpage}
\usepackage{mathtools}
\usepackage{enumerate}
\usepackage[table,xcdraw]{xcolor}

\usepackage[export]{adjustbox}
\usepackage{color}
\usepackage{csquotes}
\usepackage{array}
\newcommand\blfootnote[1]{%
  \begingroup
  \renewcommand\thefootnote{}\footnote{#1}%
  \addtocounter{footnote}{-1}%
  \endgroup
}

\usepackage{wrapfig}

\begin{document}

\title{Large-Scale Study of Perceptual Video Quality}

\author{
   Zeina Sinno, \textit{Student Member, IEEE,} and Alan Conrad Bovik, \textit{Fellow, IEEE}
}

\maketitle

\begin{abstract}The great variations of videographic skills, camera designs, compression and processing protocols, communication and bandwidth environments, and displays lead to an enormous variety of video impairments. Current no-reference (NR) video quality models are unable to handle this diversity of distortions. This is true in part because available video quality assessment databases contain very limited content, fixed resolutions, were captured using a small number of camera devices by a few videographers and have been subjected to a modest number of distortions. As such, these databases fail to adequately represent real world videos, which contain very different kinds of content obtained under highly diverse imaging conditions and are subject to authentic, complex and often commingled distortions that are difficult or impossible to simulate. 
As a result, NR video quality predictors tested on real-world video data often perform poorly. Towards advancing NR video quality prediction, we have constructed a large-scale video quality assessment database containing 585 videos of unique content, captured by a large number of users, with wide ranges of levels of complex, authentic distortions. We collected a large number of subjective video quality scores via crowdsourcing. A total of 4776 unique participants took part in the study, yielding more than 205000 opinion scores, resulting in an average of 240 recorded human opinions per video.  
We demonstrate the value of the new resource, which we call the LIVE Video Quality Challenge Database (LIVE-VQC for short), by conducting a comparison of leading NR video quality predictors on it. This study is the largest video quality assessment study ever conducted along several key dimensions: number of unique contents, capture devices, distortion types and combinations of distortions, study participants, and recorded subjective scores. The database is available for download on this link: http://live.ece.utexas.edu/research/LIVEVQC/index.html .
\end{abstract}

\blfootnote{Zeina Sinno and Alan C. Bovik are with the Department of Electrical and
Computer Engineering at The University of Texas at Austin, Austin, TX, 78712,
USA (e-mails: zeina@utexas.edu - bovik@ece.utexas.edu).}


\section{Introduction}
The goal of Video quality assesement (VQA) research is to develop video quality models that produce predictions that are in close agreement with human judgments, regardless of the video contents or the type and severities of the distortions (or the mixtures of distortions) that have corrupted the videos. Over the past decade, we have experienced a surge in the number of videos recorded, shared and watched. Sharing ``in-the-moment" experiences in the form of video has become quite popular using applications such as Instagram, Facebook, Twitter via Periscope, Snapchat, and so on.  Online videos have also revolutionized modern journalism as they enable online news stories to unfold live, and allow the viewing audience to comment on or otherwise interact with it. Over the past year, Facebook alone generated more than 100 million of video watch hours each day \cite{bergman_2017}. On YouTube, the overall durations of the videos that are uploaded daily exceeds 65 years, and more than 1 billion hours of their video content is watched each day \cite{goodrow_2017}. These numbers are continuing to rise and to reshape digital marketing, entertainment, journalism, and amateur videography. The volume of streaming video viewed online has become so large that more than 130 million people are now Netflix subscribers \cite{netflix}. Streaming videos now comprises the majority of Internet traffic today. It is no surprise that videos account for the largest portion of Internet traffic, which is expected to eclipse 82\% of all transmitted bits by 2021 \cite{cisco_2017}. These videos are captured using a very wide variety of camera devices by users having very diverse goals and expertise. 

\subsection{Conventional Laboratory VQA Databases}
It is necessary that VQA algorithms be trained and/or tested on extensive subjective video quality data sets so that it maybe asserted that they reflect or are capable of closely replicating human judgments. As a result, over the past decade numerous researchers have designed and built VQA databases.  The LIVE VQA Database \cite{seshadrinathan2010study} contains 10 pristine high-quality videos subjected to 4 distortion types: MPEG-2 compression, H.264 compression, H.264 bitstreams suffering from IP, and wireless packet losses. The resource in \cite{de2010h} offers 156 video streams suffering from H.264/AVC artifacts and wireless packet losses. The LIVE QoE Database for HTTP-based Video Streaming \cite{chen2014modeling}, studies the quality of experience of users under simulated varying channel induced distortions, and the LIVE Mobile Video Quality Database \cite{moorthy2012video} includes channel induced distortions and dynamically varying distortions, such as varying compression rates. More recent databases include the TUM databases \cite{keimel2010visual, keimel2012tum}, which target H.264/AVC distortions on a few contents (4 and 8); and the MCL-V \cite{lin2015mcl} database consists of 12 video source clips and 96 distorted videos, targeting distortions related to streaming (compression, and compression followed by scaling). The MCL video quality database contains 200 raw sequences targeting compression artifacts \cite{zhang_2016}. Most available video quality databases were conducted under highly-controlled laboratory conditions by introducing sets of graded simulated impairments onto high-quality videos. Given questions that arise regarding the realism and accuracy of representation of synthetic distortions, researchers have also conducted studies on the quality perception of authentic, real-world distortions such as distortions that occur during video capture \cite{ghadiyaram2017capture, nuutinen2016cvd2014}.

Here we aim to further advance these efforts by the construction of a new large VQA database that is more representative of real-world videos, the authentic distortions that affect them, and more typical users' opinions of their quality. By authentic distortions, we refer to  degradation that occurs in actual practice, rather than being synthesized and introduced in a controlled manner in the laboratory. The videos we gathered were captured by everyday, generally inexpert videographers using some kind of handheld camera or by smartphone. These kinds of devices now account for a preponderance of all videos being captured, watched and shared. The video capture device, level of skill of the camera operator, and the lighting conditions and other environmental factors deeply effect the visual quality of a given video. Most existing video quality assessment databases offer only very limited varieties of video content, shot by only a few users, thereby constraining the ability of learned VQA models trained on them to generalize to diverse contents, levels of videographic expertise, and shooting styles \cite{keimel2010visual, keimel2012tum, seshadrinathan2010study, chen2014modeling, lin2015mcl, ghadiyaram2017capture}. The majority of the videos in these databases were captured using high-end cameras. These were then synthetically impaired by a few controlled distortion types. However, real world videos are quite diverse and have often been subjected to  complex, nonlinear, commingled distortions that are likely impossible to accurately synthesize. To account for all these issues, we have constructed a real-world, ``in the wild", video quality assessment database which consists of 585 videos, sourced from 80 different inexpert videographers, and captured using 43 different models of 101 unique devices. The new database represents unprecedented degrees of realism, data authenticity, and relevance.

\subsection{Crowdsourced VQA Databases}
Crowdsourcing is a portmanteau of the words crowd and outsourcing. The term was first used in 2006 to describe the transfer of certain kinds of tasks from professionals to the public via the Internet.

Crowdsourcing has recently proved to be an efficient and successful method of obtaining annotations on images regarding content \cite{russakovsky2015imagenet}, image aesthetic \cite{redi2013crowdsourcing} and picture quality \cite{ghadiyaram2016massive}. 
An early effort to crowdsource video quality scores was reported in \cite{chen2010quadrant}. The authors proposed a crowdsourced framework, whereby pairwise subjective comparisons of the Quality of Experience of multimedia content (audio, images and videos) could be recorded. 

The authors of \cite{ghadiyaram2016massive} conducted a large-scale, comprehensive study of real-world picture quality and showed that their results were quite consistent with the results of subjective studies conducted in a laboratory.

This success of latter study \cite{ghadiyaram2016massive} has inspired our work here, with a goal to build a large, diverse and representative video database on which we crowdsourced a large-scale subjective video quality study. We encountered many difficulties along the way, many of which were significantly more challenging than in the previous picture quality study \cite{ghadiyaram2016massive}. The issues encountered from simple participant problems (distraction, reliability and a imperfect training \cite{seufert2016one}), to more serious issues such as variations in display quality, size and resolution, to very difficult problems involving display hardware speed and bandwidth conditions on the participant's side.
We carefully designed a framework in Amazon Mechanical Turk (AMT) to crowdsource the quality scores while accounting for these numerous factors, including low bandwidth issues which could result in video stalls, which are very annoying during viewing and can heavily impact the experienced video quality.

Previous crowdsourced video quality studies have not addressed the latter very important concern. For example, in the study in \cite{keimel2012qualitycrowd}, the participants were allowed to use either Adobe Flash Player or HTML5 to display videos, depending on the compatibility of their browser. However, in their methodology, no assurance could be made that the videos would fully preload before viewing, hence there was no control over occurrences of frame freezes or stalls, or even to record such instances on the participants' end, as Flash Player does not have this option, and some browsers disable this option for HTML5 video element. When the videos are not preloaded and are streamed instead, interruptions and stalls are often introduced, whereas the study in \cite{keimel2012qualitycrowd} did not report any effort to record whether such events took place. The early QoE crowdsource framework \cite{chen2010quadrant} also did not report any accounting of this important factor.  

A significant and recent crowdsourced VQA database was reported in \cite{hosu2017konstanz}, providing an important new resource to the video quality community. In this study, a subject was asked to rate any number of videos within the range 10 to 550 videos. We note that viewing as many as videos as the upper end of this range is likely to produce fatigue, which affects performance. Generally, it is advisable to restrict the number of watched videos per session so that the session time does not exceed 30-40 mins including training, to reduce fatigue or loss of focus \cite{rec2009bt}. Also, we observed that the study participants in \cite{hosu2017konstanz} were allowed to zoom in or out while viewing, which can introduce artifacts on the videos, and can lead to interruptions and stalls. Scaling a video up or down is computationally expensive. Under these conditions, stalls could occur even if a video was fully preloaded into memory before being played. Downscaling and upscaling artifacts, and video stalls are factors that significantly impact the perceived video quality.  Detecting whether stalls occurred is also critical. It appears that the authors of the study did not account in any way for stalls, which is highly questionable, since stalls can deeply impact reported subjective quality. 
These types of issues underline the difficulty of online video quality studies, and the need for careful design of the user interface, monitoring of the subjects, and the overall supporting pipeline used to execute a large-scale study.

A few other video crowdsourcing methods have been reported, at much smaller scales without addressing the difficult technical issues \cite{hossfeld2014best, figuerola2013assessing, shahid2014crowdsourcing,chen2015qos, rainer2014quality} described in the preceding. 
\begin{figure*}[!h]
\centering
\captionsetup[subfigure]{labelformat=empty}
\begin{tabular}{@{}c@{}c@{}c@{}c@{}c@{}c@{}}
& ~ & ~ & ~ & ~  & ~\\
\begin{minipage}[c]{0.0cm}\vspace{1.80cm}\end{minipage}&\subfloat[]{\includegraphics[height= 2.8cm, width=5.2cm]{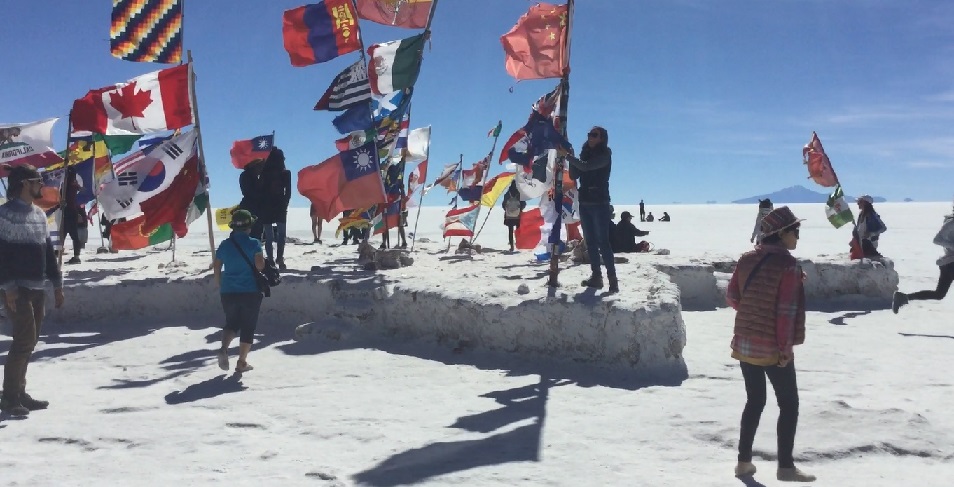}}\hspace*{2mm}
&\subfloat[]{\includegraphics[height= 2.8cm, width=5.2cm]{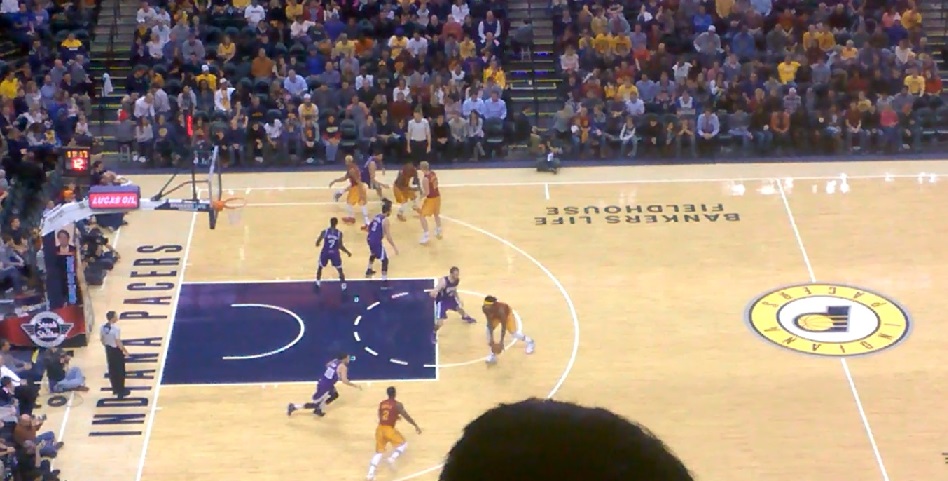}}\hspace*{2mm}
&\subfloat[]{\includegraphics[height= 2.8cm, width=5.2cm]{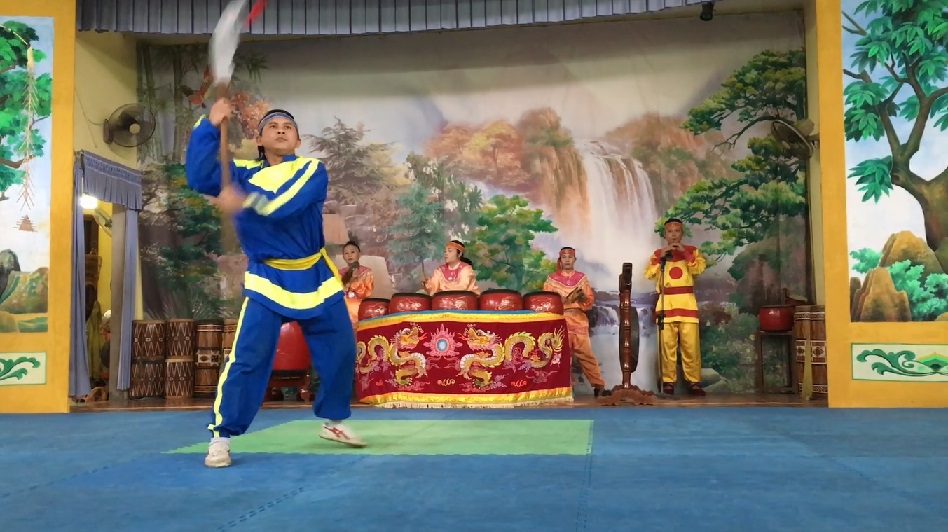}}\vspace*{-15mm}\\
\end{tabular}
\begin{tabular}{@{}c@{}c@{}c@{}c@{}c@{}c@{}}
& ~ & ~ & ~ & ~  & ~\\
\begin{minipage}[c]{0.0cm}\vspace{1.8cm}\end{minipage}&\subfloat[]{\includegraphics[height=  4.2cm, width=3.01cm]{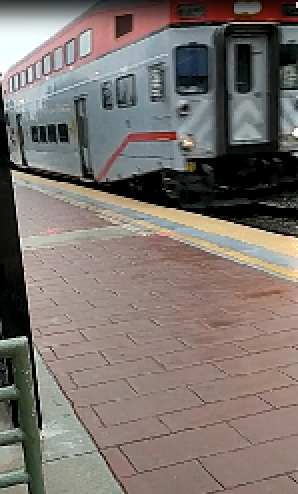}}\hspace*{2.0mm}
&\subfloat[]{\includegraphics[height=  4.2cm, width=3.01cm]{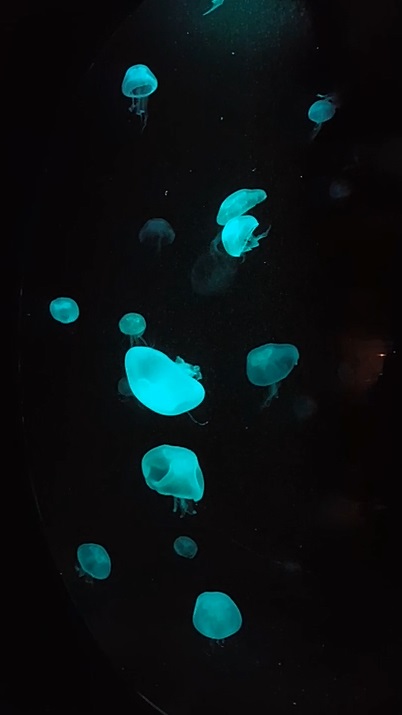}}\hspace*{2.0mm}
&\subfloat[]{\includegraphics[height=  4.2cm, width=3.01cm]{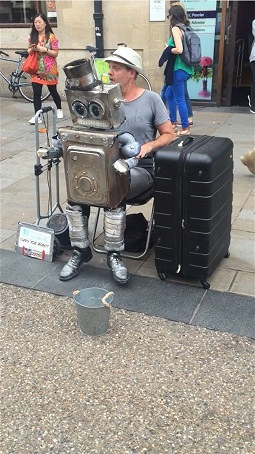}}\hspace*{2.0mm}
&\subfloat[]{\includegraphics[height=  4.2cm, width=3.01cm]{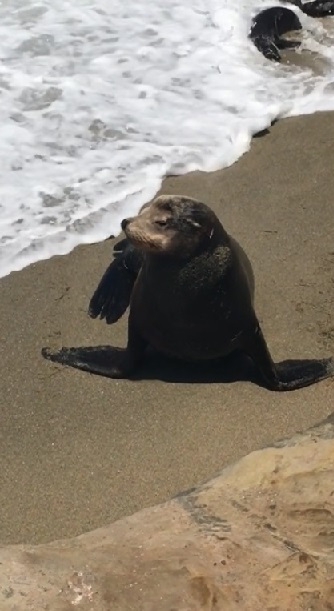}}\hspace*{2.0mm}
&\subfloat[]{\includegraphics[height=  4.2cm, width=3.01cm]{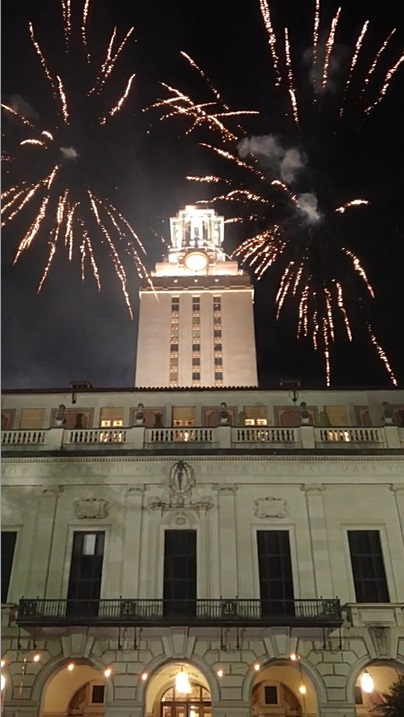}}\vspace*{-15mm}\\
\end{tabular}
\begin{tabular}{@{}c@{}c@{}c@{}c@{}c@{}c@{}}
& ~ & ~ & ~ & ~  & ~\\
\begin{minipage}[c]{0.0cm}\vspace{1.8cm}\end{minipage}&\subfloat[]{\includegraphics[height= 2.8cm, width=5.2cm]{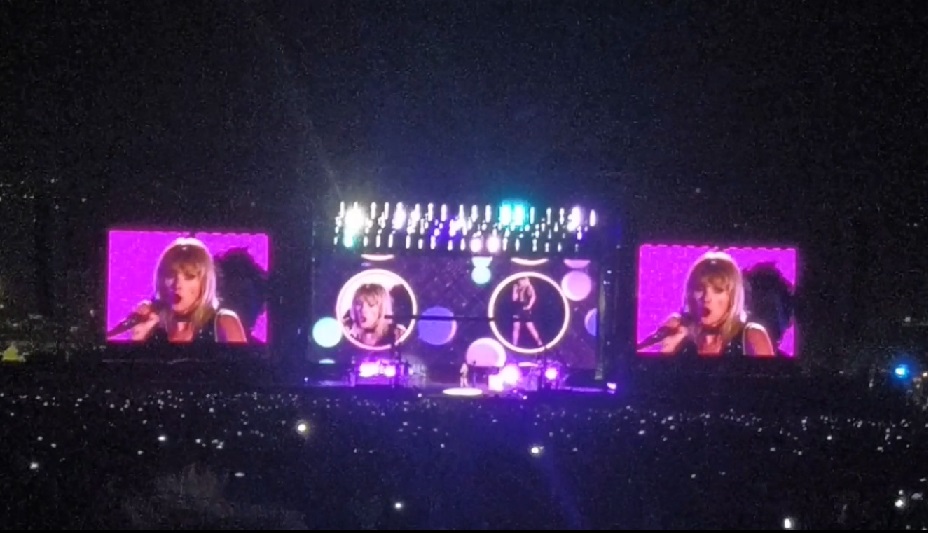}}\hspace*{2mm}
&\subfloat[]{\includegraphics[height= 2.8cm, width=5.2cm]{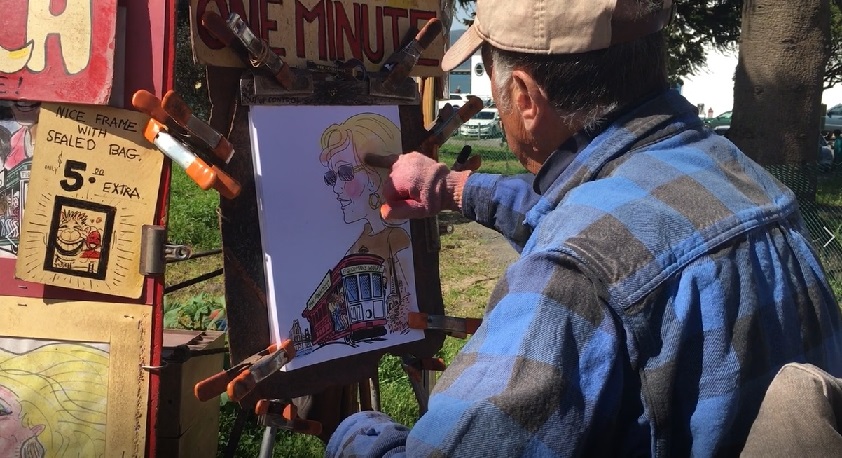}}\hspace*{2mm}
&\subfloat[]{\includegraphics[height= 2.8cm, width=5.2cm]{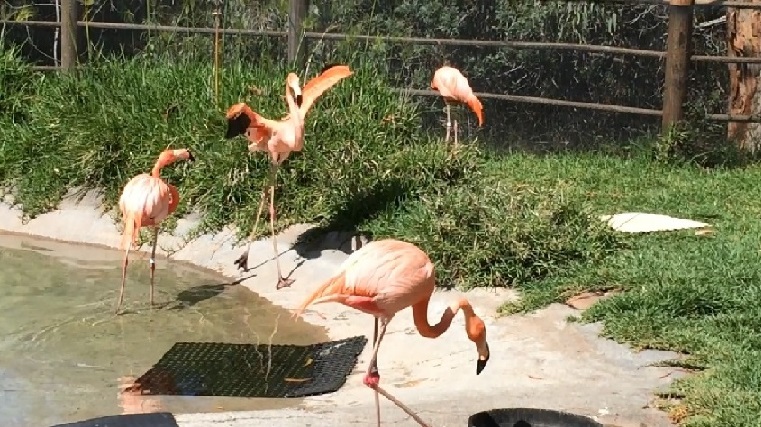}}\vspace*{-15mm}\\
\end{tabular}
\begin{tabular}{@{}c@{}c@{}c@{}c@{}c@{}c@{}}
& ~ & ~ & ~ & ~  & ~\\
\begin{minipage}[c]{0.0cm}\vspace{0.7cm}\end{minipage}&\subfloat[]{\includegraphics[height=  4.2cm, width=3.01cm]{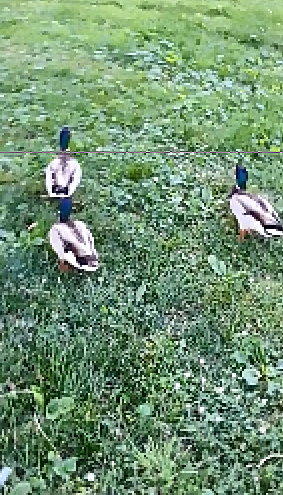}}\hspace*{2.0mm}
&\subfloat[]{\includegraphics[height=  4.2cm, width=3.01cm]{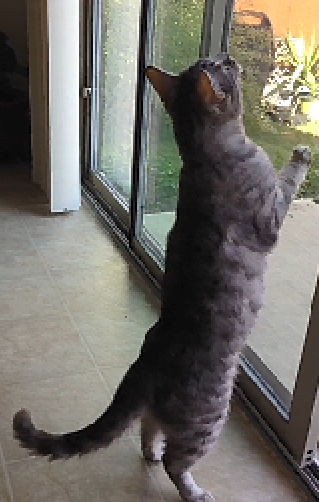}}\hspace*{2.0mm}
&\subfloat[]{\includegraphics[height=  4.2cm, width=3.01cm]{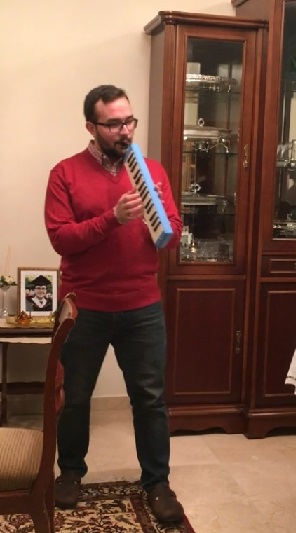}}\hspace*{2.0mm}
&\subfloat[]{\includegraphics[height=  4.2cm, width=3.01cm]{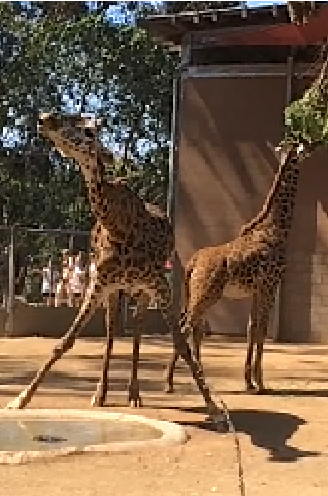}}\hspace*{2.0mm}
&\subfloat[]{\includegraphics[height=  4.2cm, width=3.01cm]{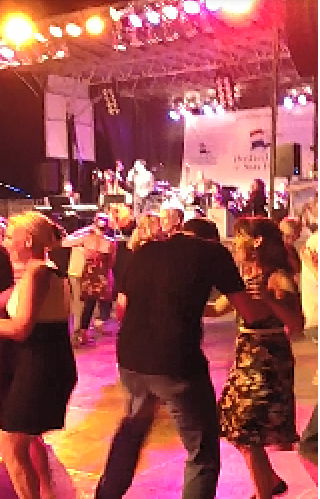}}\vspace*{-5mm}\\
\end{tabular}

\caption{Screenshots of frames from some of those presented during the study.}
\label{fig:SampleVideos}

\end{figure*}

The contributions of the paper are summarized as follows:
\begin{enumerate}
\item{A new robust framework for conducting crowdsourced video quality studies and for collecting video quality scores in AMT.}
\item{A new VQA database, that we call the LIVE Video Quality Challenge (LIVE-VQC), which contains contains 585 videos of unique content and associated MOS, captured by 80 different users and 101 devices. The videos are afflicted by authentic distortions like those that occur in the real world. We are making this database available to the research community at no charge. \cite{linkVQC}.}
\end{enumerate}

In the following sections, we present the results of a large-scale crowdsourced video study, where we collected more than 205000 opinion scores on 585 diverse videos containing complex authentic distortions. We also evaluate the performances of prominent blind VQA algorithms on the new database. The rest of the paper is organized as follows: we present the details of the video database in Section \ref{sec:Database}, the experimental methods in Section \ref{sec:Methodology}, the experimental methods of post-processing of the subjective scores and the outcomes of our study in Section \ref{sec:RawScores}, and the algorithm comparisons in Section \ref{Algorithms}. The paper concludes with future thoughts in Section \ref{Conclusion}.

 \section{LIVE Video Quality Challenge (LIVE-VQC) Database}
 \label{sec:Database}
While considerable effort has been applied to the VQA problem for high-end streaming video (e.g., Internet television), much less work has been done on videos captured by mobile and digital cameras by casual users. Our objective was to create a resource to support research on this very large-scale, consequential topic. Our specific aim is to offer a large, high-quality dataset of authentically captured and distorted videos, and a large corpus of science-quality psychometric video quality scores.

\subsection{Content Collection}

Our data was collected with the assistance of 80 largely na\"{i}ve mobile camera users from highly diverse age groups, gender, and social, cultural and geographic diversity. We requested the collaborators to upload their videos just as captured, without any processing (for example by video processing `apps' like Instagram or Snapchat). Only videos having durations of at least 10 seconds were accepted. No instructions regarding the content or capture style was provided, other than it reflect their normal use.

Most of the video contributors were volunteers, including acquaintances of LIVE members; i.e, from family, friends, friends of friends, and so on, from around the world, while the rest ($\sim$18\%) were students solicited from the undergraduate and graduate population at The University of Texas at Austin. The number of videos provided by each contributing videographer varied but none contributed by more than 9\% of the video set, to ensure diversity of method, content and style.  The contributors spanned a wide age range (11 to 65 years old), and were divided about evenly by gender. The content was shot on all the populated continents and in many countries, including Australia, U.S.A., Mexico, Peru, Panama, Colombia, Bolivia, India, Malaysia, Vietnam, China, South Korea, Germany, Norway, Switzerland, Poland, Sweden, U.K., Portugal, Turkey, Lebanon, the United Arab Emirates, Oman, Tunisia, Egypt and more. More than 1000 videos were gathered, cropped to 10 seconds while seeking to preserve `story' continuity, culled to remove redundant content captured by a same user and videos with disturbing content (e.g. a scene of surgery). After this cleaning, we were left with 585 videos.

As exemplified in Fig. \ref{fig:SampleVideos}, the obtained video content is quite diverse, and includes scenes of sports games, music concerts, nature, various human activities (parades, dancers, street artists, cowboys etc.), and much more. The scenes were captured under different lighting conditions (different times of day and night), and include both indoor and outdoor scenes. Widely diverse levels of motion (camera motion and in-frame motion) are present and often contibute to complex, space variant distortions. While it is very difficult to categorize real-world picture and video distortions with any precision, owing to their intrinsic mutability, their tendency to commingle, many distortions have been observed including, for example, poor exposures, and a variety of motion blurs, haziness, various imperfect color representations, low-light effects including blur and graininess, resolution and compression artifacts, diverse defocus blurs, complicated combinations of all of these, and much more. The interactions of multiple artifacts also give rise to very complex, difficult to describe composite impairments. Often visible distortions appear, disappear, or otherwise morph during a video, as for example, temporary autofocus blurs, exposure adjustments, and changes in lighting. As such, we made no attempt to supply distortion labels to the videos.

\subsection{Capture Devices}
\begin{figure}[H]\vspace{-0.90cm}
\centering
\captionsetup[subfigure]{labelformat=empty}
\begin{tabular}{@{}c@{}c@{}c@{}c@{}c@{}c@{}}
& ~ & ~ & ~ & ~  & ~\\
\begin{minipage}[c]{0.0cm}\vspace{-0.20cm}\end{minipage}&\subfloat[]{\includegraphics[width=8.0cm,frame]{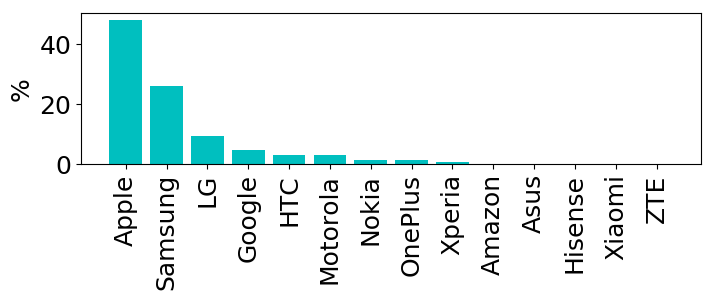}}\hspace*{0mm}\\
\end{tabular}

\caption{Distribution of viewed videos grouped by device brand.}
\label{fig:StatsVideos1}
\end{figure}

A taxonomy of the mobile devices used to capture the videos is given in Table \ref{TableDevices}. Unsurprisingly, the majority of these were smartphones. A total of 101 different devices were deployed (some users provided videos captured by multiple devices), including 43 different models. The commercial releases of the devices varied between 2009 and 2017, although most of the videos were captured using devices that were released after 2015 and beyond.
\begin{table}[!h]
\centering
\caption{Number of videos captured by each type of camera devices.}
\label{TableDevices}
\begin{tabular}{|c|c|c|}
\hline
Make     & Model          & Number of Videos \\ \hline
Amazon   & Fire HDX       & 1                \\ \hline
Apple    & Ipad Pro       & 2                \\ \hline
Apple    & Iphone 3GS     & 1                \\ \hline
Apple    & Iphone 4       & 14               \\ \hline
Apple    & Iphone 4S      & 2                \\ \hline
Apple    & Iphone 5       & 25               \\ \hline
Apple    & Iphone 5s      & 49               \\ \hline
Apple    & Iphone 6       & 48               \\ \hline
Apple    & Iphone 6s      & 107              \\ \hline
Apple    & Iphone 6s plus & 5                \\ \hline
Apple    & Iphone 7       & 17               \\ \hline
Apple    & Iphone 7 plus  & 3                \\ \hline
Apple    & Ipod touch     & 8                \\ \hline
Asus     & Zenfone Max    & 1                \\ \hline
Google   & Pixel          & 7                \\ \hline
Google   & Pixel XL       & 20               \\ \hline
Hisense  & S1             & 1                \\ \hline
HTC      & 10             & 13               \\ \hline
HTC      & M8             & 5                \\ \hline
Huawei   & Nexus 6P       & 10               \\ \hline
LG       & G3             & 3                \\ \hline
LG       & G4             & 2                \\ \hline
LG       & Nexus 5        & 50               \\ \hline
Motorola & E4             & 1                \\ \hline
Motorola & Moto G 4G      & 3                \\ \hline
Motorola & Moto G4+       & 1                \\ \hline
Motorola & Moto Z Force   & 12               \\ \hline
Nokia    & Lumia 635      & 5                \\ \hline
Nokia    & Lumia 720      & 3                \\ \hline
OnePlus  & 2              & 4                \\ \hline
OnePlus  & 3              & 4                \\ \hline
Samsung  & Core Prime     & 5                \\ \hline
Samsung  & Galaxy Mega    & 1                \\ \hline
Samsung  & Galaxy Note 2  & 21               \\ \hline
Samsung  & Galaxy Note 3  & 5                \\ \hline
Samsung  & Galaxy Note 5  & 72               \\ \hline
Samsung  & Galaxy S3      & 4                \\ \hline
Samsung  & Galaxy S5      & 25               \\ \hline
Samsung  & Galaxy S6      & 14               \\ \hline
Samsung  & Galaxy S8      & 6                \\ \hline
Xiaomi   & MI3            & 1                \\ \hline
Xperia   & 3 Compact      & 3                \\ \hline
ZTE      & Axon 7         & 1                \\ \hline
\end{tabular}
\end{table}
\\Figure \ref{fig:StatsVideos1} depicts the distribution of the viewed videos grouped by brand. As expected \cite{Gartner} the majority of these videos ($\sim$ 74\%) were captured using commercially dominant Apple and Samsung devices.

\subsection{Video Orientations and Resolutions}
We imposed no restrictions on the orientation of the camera device during capture (or after), and 23.2\% of the videos in the database were taken in portrait mode and the other 76.2\% in landscape mode. The majority of the videos shot in portrait and are of high resolutions (1080$\times$1920 and 3840$\times$2160) which cannot be fully displayed by most available displays without downscaling them. The median display configuration in use today appears to be 1280$\times$720 \cite{Resolutions}. To ensure compatibility, all portrait videos of resolutions 1080$\times$1920, 2160$\times$3840, and 720$\times$1080 were downscaled using bicubic interpolation to 404$\times$720, so that they could be displayed at the native display resolutions of all subjects accepted to participate in the study.

Among the videos in landscape mode, many were of resolutions that cannot be displayed by viewers (those that were 1920$\times$1080 and 3840$\times$2160). At the time the study was conducted, it was estimated that only between 10-20\% of global web viewers possessed high resolution displays equal to or exceeding 1920$\times$1080 \cite{Resolutions}. As a way of accessing both high and low resolution display groups, we decided to downscale a portion of the large resolution videos to 1280$\times$720 to better distribute the scoring tasks, since we expected relatively few participants to be capable of viewing high resolutions. Thus, 110 videos randomly selected videos were maintained at resolution of 1920$\times$1080, while the remaining 1920$\times$1080 and higher resolution videos were downsampled to 1280$\times$720 using bicubic interpolation. We ended up with 18 different resolutions in our database, as summarized in Table \ref{VideoResolutions}.  
\begin{table}[!h]
\centering
\caption{Video resolutions in the database}
\label{VideoResolutions}
\centering
\resizebox{\columnwidth}{!}{%
\begin{tabular}{|c|c|c|c|c|c|}
\hline
1920$\times$1080 & 1280$\times$720 & 960$\times$540 & 800$\times$450 & 480$\times$640 & 640$\times$480 \\ \hline
404$\times$720   & 360$\times$640  & 640$\times$360 & 352$\times$640 & 640$\times$352 & 320$\times$568 \\ \hline
568$\times$320   & 360$\times$480  & 480$\times$360 & 272$\times$480 & 240$\times$320 & 320$\times$240 \\ \hline
\end{tabular}}
\end{table}

\begin{figure}[!hbt]\vspace*{-1cm}
\centering
\captionsetup[subfigure]{labelformat=empty}
\begin{tabular}{@{}c@{}c@{}c@{}c@{}c@{}c@{}}
& ~ & ~ & ~ & ~  & ~\\
\begin{minipage}[c]{0.0cm}\vspace{-0.20cm}\end{minipage}&\subfloat[]{\includegraphics[width=6.20cm,frame]{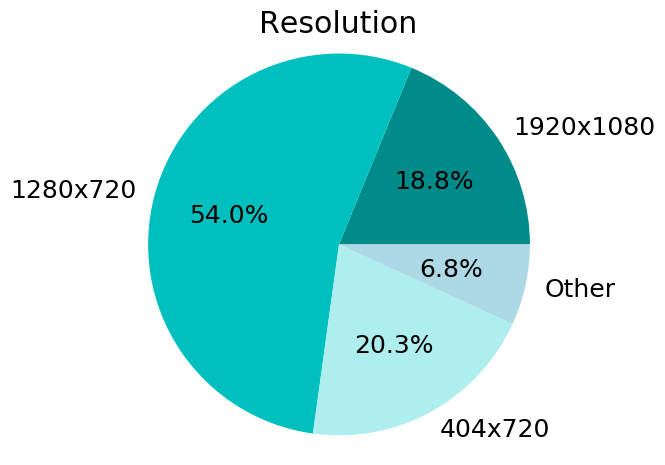}}\hspace*{0mm}\\
\end{tabular}

\caption{Distribution of the resolutions viewed by the AMT workers.}
\label{fig:StatsVideos2}
\end{figure}

The predominant resolutions were 1920$\times$1080, 1280$\times$720 and 404$\times$720, which together accounted for 93.2\% of the total, as shown in Fig. \ref{fig:StatsVideos2}. The other resolutions combined accounted for 6.8\% of the database.

We managed the videos shown to each worker by first detecting their display resolution by executing code in the background. If their display resolution was at least 1920$\times$1080, then half of the videos they evaluated would have this resolution, whilst the rest were randomly selected from the rest of the database. All of the other participants viewed randomly selected videos having resolutions less than 1920$\times$1080.

\section{Testing Methodology}
\label{sec:Methodology}
Given that real users nearly always view single videos, rather than side-by-side pairs, and since, in every case we have only a single, authentically distorted version of each content, we deployed a single stimulus presentation protocol. Since the study is crowdsourced and conducted in the wild, we could not apply many ITU standard recommendations (e.g. \cite{rec2009bt}) when conducting the subjective studies \cite{sinno2018icip}. We did, however abide by agreed-on principles regarding timing, stimulus presentation, and subject training, as detailed in the following.
\begin{figure*}[!htbp] 
\begin{minipage}[b]{1.0\linewidth}
  \centering
  \captionsetup{justification=centering}
   \centerline{\includegraphics[width=16cm]{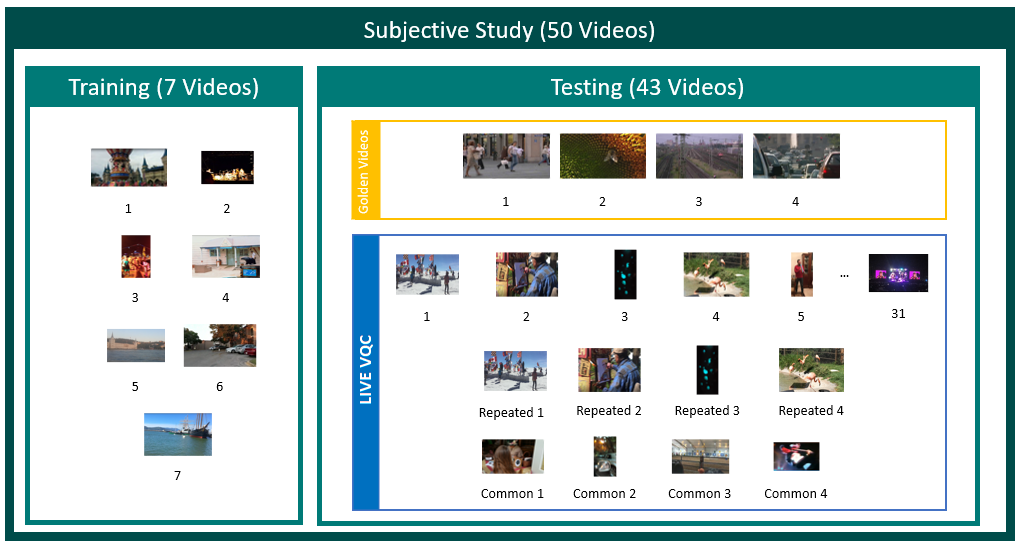}}
\end{minipage}
\caption{Chart showing the categories of videos seen by a subject viewing session.}
\label{fig:VideosViewed}
\end{figure*}
\subsection{Participation Requirements}
We first list the participation requirements, then explain each in detail. To be eligible to participate, the worker should:
\begin{enumerate}[1)]
\item Have an AMT reliability score above 90\% (reliability constraint).\label{Cond0} 
\item Have not participated previously in the study (unique worker constraint). \label{Cond1} 
  \item Use a non-mobile display device, \textit{viz.} desktop and laptops are allowed, while mobile phones and tablets are not (display constraint). \label{Cond2}
  \item Use a display having a minimum resolution of 1280$\times$720 (resolution constraint). \label{Cond3}
\item Use a recently updated supported browser. The study supports Google Chrome, Safari, Mozilla Firefox, and Opera. Internet Explorer, Microsoft Edge and other browsers were not allowed (browser constraint). \label{Cond4}
\item Have a good Internet capacity (connectivity constraint). \label{Cond6}
\item Use a device with adequate computational power (hardware constraint). \label{Cond7}
\end{enumerate}
Explanations supporting these choices are as follows:
\subsubsection{Reliability constraint}
AMT records how many jobs each worker has completed, and how many jobs were accepted, to determine the acceptance ratio, known as the reliability score. Because of the subtlety of many distortions and to better ensure subject assiduity, we only allowed workers having an acceptance rates exceeding 90\% to participate.
\subsubsection{Unique worker constraint}
We imposed this condition to avoid any judgment biases that might arise if workers rated videos more than once.
\subsubsection{Device constraint}
We enforced this condition for two reasons. First, mobile browsers do not support preloading videos, which is a major concern. Second, it is not possible to control the resolutions of videos displayed on mobile browsers, since they must be played using a native player on the device where are upscaled or downscaled, then played in full screen mode, whereby additional, unknown artifacts are introduced. 

\subsubsection{Resolution constraint}
We required the worker display resolutions to be at least 1280$\times$720 (720p) as discussed earlier.
\subsubsection{Browser constraint}
As of early 2018, Internet Explorer and Microsoft Edge do not support video preloading in HTML5. For this reason, we did not allow users of those browsers to take part of the study. Google Chrome, Safari, Mozilla Firefox, and Opera support this option starting at a certain version. We verified that the browser (and the version) used by each worker was compatible with the HTML5 video  preloading attribute. We verified that each session could proceed with smooth preloading, thereby eliminating the possibility of bandwidth-induced stalling (rebuffering) events. 

\subsubsection{Connectivity constraint} Poor Internet connectivity or slow bandwidths can cause annoying delays as the videos are loading leading to possible frustration and loss of focus on the part of the subjects. Under extremely poor bandwidth conditions, it can also lead to timeouts in the connection established between the server where the videos are stored and the worker's side. Internet bandwidth is stochastic and unpredictable as it changes over time. In rare cases, a timeout can emerge at the users' side, with good bandwidth conditions. For example, a new device may join the network and initiate a large download. In such a case, a sudden drop in the bandwidth could be experienced. 
To minimize these problems, we tracked the loading progress of each video and acted accordingly. Each video was requested at least 30 seconds before it was needed by the subject. If the connection was not successfully established the first time, a second attempt was made 10 seconds later, and a third 10 seconds after that. If the connection again failed, the session was terminated and the worker was informed. Once a connection was successfully established and the loading commenced, if it was detected that the loading progress halted for a certain interval of time, the connection with the server was terminated and a new one established. This was allowed to occur only once. As a global constraint, the duration of each study session was not allowed to exceed 30 minutes. This helped to filter out corner cases where connections were successfully established but the loading progress was very slow. We also implemented two temporal checkpoints to track the progress of each worker. After a third of the session videos were viewed, if it was detected that more than 10 minutes (one third of the allowed time) had elapsed, then a warning message was displayed informing the worker that they might not be able to complete the test on time. The second checkpoint occurred after two thirds of the content had been viewed, warning them if 20 minutes had passed. We encouraged the workers (as part of the training process) to close any other windows or tabs open in the background before commencing the study to avoid draining their bandwidth. They were reminded of this again if their progress was slow or if they were experiencing large delays. Before launching the study, we extensively tested the framework under highly diverse bandwidth conditions and scenarios to ensure its efficacy.
\begin{figure*}[!h]
\begin{minipage}[b]{1.0\linewidth}
  \centering
  \captionsetup{justification=centering}
   \centerline{\includegraphics[width=17.00cm, frame]{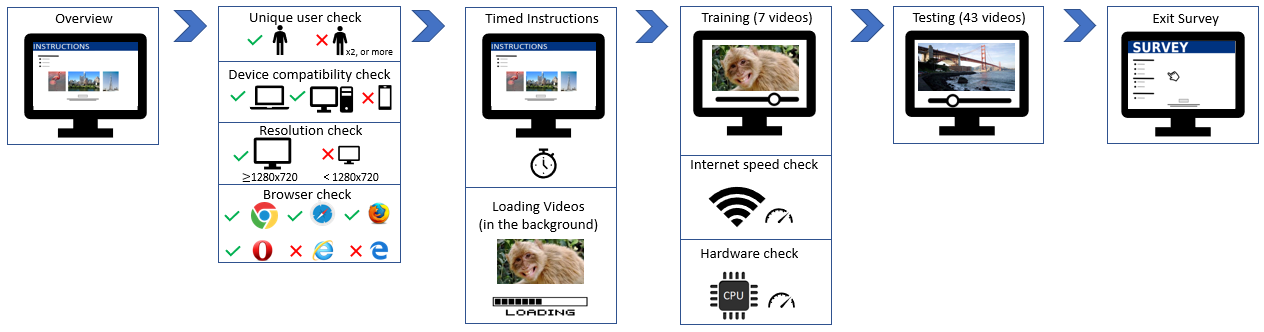}}
\end{minipage}
\caption{Subjective study workflow.}
\label{fig:FlowChart}
\end{figure*}
\subsubsection{Hardware constraint} Slow hardware or poor computational power can lead to ``frame freeze'' stalls while a video is played. To minimize the frequency of these occurrences, we encouraged each worker (via the training instructions) to close any programs running in the background, and if they were using a laptop, to ensure that it was plugged into an outlet to further promote better performance.

\subsection{Viewed Content}
During a single session, each subject viewed a total of 50 videos: 7 training and the remaining 43 during the rating process (Fig. \ref{fig:VideosViewed}).
The videos displayed during the training process were selected to broadly span the ranges of video quality containing within the database, and were of mixed resolutions, to prepare the subjects for their later viewing. The training varied slightly with the detected sizes of the workers' displays. Those viewers using displays of resolution of 1920$\times$1080 or higher were presented with two videos matching their display, along with a mixture of videos of smaller resolutions (1280$\times$720 and less). Those subjects having display resolutions lower than 1920$\times$1080 were presented with videos of mixed resolutions no higher than 1280$\times$720.

The 43 videos viewed during the judgment (test) phase included:
\begin{itemize}
\item 4 distorted videos drawn from the LIVE Video Quality Assessment Database \cite{seshadrinathan2010study}, which we will refer to as the ``golden videos.'' These videos were previously rated by human viewers in the tightly controlled study  \cite{seshadrinathan2010study}, and are used, along with the prior subjective scores from  \cite{seshadrinathan2010study}, as a control to validate the subjects' ratings.
\item 31 videos randomly selected from the new distorted video database. If the worker had a display resolution no less than 1920$\times$1080, then 18 videos were drawn from the pool of videos having a resolution of 1920$\times$1080, and the remaining 13 videos selected from the other, lower resolution videos.
\item 4 videos randomly selected from the same pool of 31 videos as above, but repeated at relatively displayed moments as a control.
\item 4  videos selected from the database were viewed and rated by all of the workers.
\end{itemize}
The 43 videos were placed in re-randomized order for each subject.

\subsection{Experimental Flow}
Each subjective study session followed the workflow depicted in Fig. \ref{fig:FlowChart}. We now describe each step in detail.
\subsubsection*{\textbf{Step 1: Overview}}
Once a worker with a reliability score exceeding 90\% selected our study to preview it, s/he was prompted to an overview page describing the task, the requirements to participate (conditions \ref{Cond1}-\ref{Cond7}), the instructions on how to rate a video, and a few example videos to give them a clearer sense of the task. The worker was instructed to rate the videos based on how well s/he believes the presented video quality compares to an ideal, or best possible video of the same content. Several example videos were then played to demonstrate exemplars of some of the video distortions such as under exposure, stalls, shakes, blur and poor color representation. The worker was informed that other types of distortions exist and would be seen, so the worker would not supply ratings based only on the exemplar types of distortions, but would instead rate all distortions. 

\subsubsection*{\textbf{Step 2: Eligibility check}}
If the user accepted to work on the `hit', and it was determined whether s/he did not previously participate in the study, and that s/he met conditions \ref{Cond1})-\ref{Cond4}) above. If the worker did not meet any of those conditions, a message was displayed indicating which condition was unmet, and that s/he was kindly requested to return the hit. If it was the case that any of conditions \ref{Cond2})-\ref{Cond4}) was unmet, then the displayed message invited the worker to try working on the hit again, but using another device/ browser etc., depending on which condition was not met. During this step, the browser zoom level was adjusted to 100\% to prevent any downscaling or upscaling artifacts from occurring when the videos are played.

\subsubsection*{\textbf{Step 3: Timed Instructions}}
If the worker was able to proceed, the instruction page was displayed again, with a countdown timer of one minute. Once the countdown timer reached zero, a proceed button would appear at the bottom of the page, thereby allowing the worker to move forward. The instructions were repeated because while the study was in progress (at the end of 2017), AMT was in the process of migrating towards a new user interface that allowed the users to accept a `hit' without first previewing it. Additionally, some workers used scripts while working on AMT, such as Microworkers and Tampermonkey, which would auto-accept hits on behalf of the workers when posted. Hence, some of the workers would not have had the opportunity to read the instructions if they were not repeated. While the instructions were being repeated, the first three videos began loading in the background, and the videos that were to be displayed during the testing phase were determined.

\begin{figure*}[!htb]
\centering
\captionsetup[subfigure]{labelformat=empty}
\begin{tabular}{@{}c@{}c@{}c@{}c@{}c@{}c@{}}
& ~ & ~ & ~ & ~  & ~\\
\begin{minipage}[c]{0.0cm}\vspace{-0.0cm}\end{minipage}&\subfloat[a) Step 1: viewing a video.]{\includegraphics[width=13.5cm,frame]{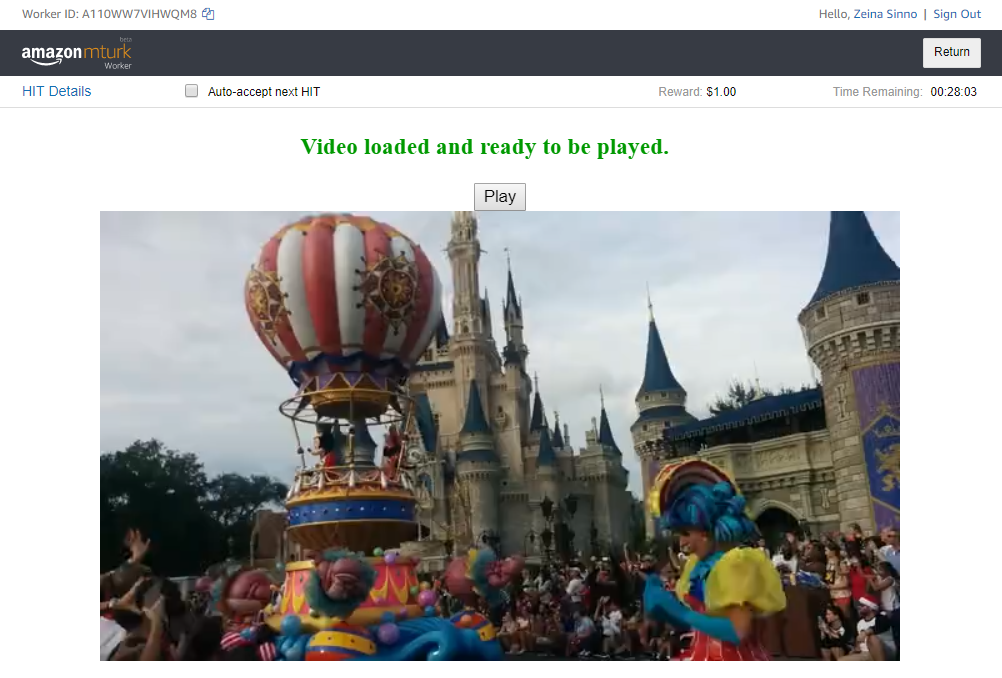}}\vspace*{-2mm}\hspace*{0mm}\\
&\subfloat[b) Step 2: rating the video ]{\includegraphics[width=13.5cm,frame]{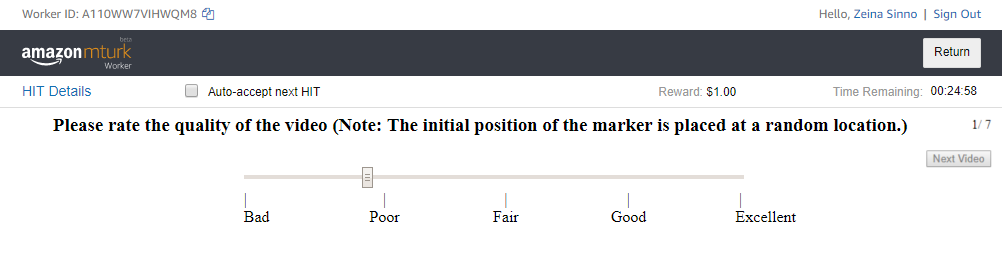}}\vspace*{0mm}\\
\end{tabular}

\caption{Screenshot of the interface used to rate the videos.}
\label{fig:RatingVideo}
\end{figure*}

\subsubsection*{\textbf{Step 4: Training}}
Once a subject clicked on the \textit{Proceed} button, a message was displayed indicating that the training phase was about to start. This phase consisted of 7 videos. 
A screenshot of the interface featuring each video to be rated (during both the training and test phase) is shown in Fig. \ref{fig:RatingVideo}. As shown, the video controls were hidden and inaccessible, to prevent less dedicated workers from pausing, replaying or skipping the videos. Before a video was fully loaded, a message was displayed showing the loading progress. Once the video was fully loaded, a message informed the user that ``Video loaded and ready to be played." At this point, the zoom level was checked to determine whether it was at 100\%, and was then adjusted if need to be. The video was then played in entirety (while being muted) if the page was displayed in full screen mode. Otherwise, a message was displayed directing the worker to adjust it and to again press the play button afterwards. This had the additional benefit of reducing worker multitasking, which could distract attention from the video task.

Once each video finished playing, it disappeared, revealing the rating interface shown in Fig. \ref{fig:RatingVideo}(b). A continuous bar allowed the workers to rate the quality of the videos, where a Likert-like scale with 5 marks; Bad, Poor, Fair, Good, and Excellent was provided to generally guide their ratings. The initial position of the cursor was randomized. This was mentioned in the instructions and was also indicated in a note above the rating bar. Once each video finished playing, the user moved the cursor to the appropriate position. The user was not allowed to proceed to the next video unless s/he moved the position of the cursor. Once a change in the cursor location was detected, the \textit{Next Video} button became clickeable. Once clicked, the worker moved to a new page, with a new video to be rated and the process continued until the last video had been rated.

A number of processes were ongoing as the user was viewing/rating rating each video. For example, the following videos to be rated next start would begin loading in the background, as described in the previous section. 
During the training process, the play duration of each video was measured to assess the workers' play capability. There are many ways that stalls could occur while a video is playing. If a worker's hardware CPU was slow, if other programs were running in the background (CPU is busy) or if the Internet connection was poor, then stalls or frame freezes could (and did) occur. Required background tasks (such as loading the videos to be played next) added processing overhead, while slower Internet bandwidths required increased processing overhead, further impacting foreground performance. During the training process, 7 videos of 10 seconds duration each were played. Importantly, the workers were not able to proceed further if it took more than 15 seconds to play any of the 7 videos or if any 3 of the 7 videos each required more than 12 seconds to play. Adopting this strategy guaranteed that most of the training videos were played smoothly, and also allowed us to eliminate workers who were unlikely be able to successfully complete the `hit.'

\subsubsection*{\textbf{Step 5: Testing}}
After the training phase was completed, a message was displayed indicating that the video rating phase was about to begin. The testing phase was very similar to the training phase; the videos were displayed, controlled and rated in the same way. However, the testing phase required 43 videos to be rated, instead of 7.

Once a third of the study was completed, (10 testing videos rated), if the progress of the worker was sufficient, then the following message was displayed: ``You have completed one third of the overall study! Good Job :-) Keep up the Good Work!'' As shown in \cite{vondrick2013efficiently}, providing workers with motivational feedback can encourage them to provide work of better quality.
If the progress of the worker was slow ($>$10 minutes had passed), the following message was displayed ``You have completed one third of the overall study but your progress is slow. Are the videos taking too long to load? If so, make sure to close any programs running in the background." A similar message was displayed after two thirds of the study was completed.
\subsubsection*{\textbf{Step 6: Exit Survey}}
Once the worker finished rating all of the videos, s/he is directed to the exit survey so that information regarding the following factors could be collected:
\begin{itemize}
\item the display,
\item viewing distance,
\item gender and age of the worker,
\item country where the task study was undertaken,
\item whether the worker needs corrective lenses, and if so, if s/he wore them.
\end{itemize}
The subjects were also asked whether they had any additional comments or questions. At the same time, information was automatically collected regarding the display resolution.
\subsection{Human Subjects}
\subsubsection{Demographic Information}
\begin{figure}[!h]\vspace*{-0.7cm}
 \centering
\captionsetup{justification=centering}
\begin{tabular}{@{}c@{}c@{}c@{}c@{}c@{}c@{}}
& ~ & ~ & ~\\
\begin{minipage}[c]{0.0cm}\vspace{4.00cm}\end{minipage}&\subfloat[]{\includegraphics[width=8.8cm,frame]{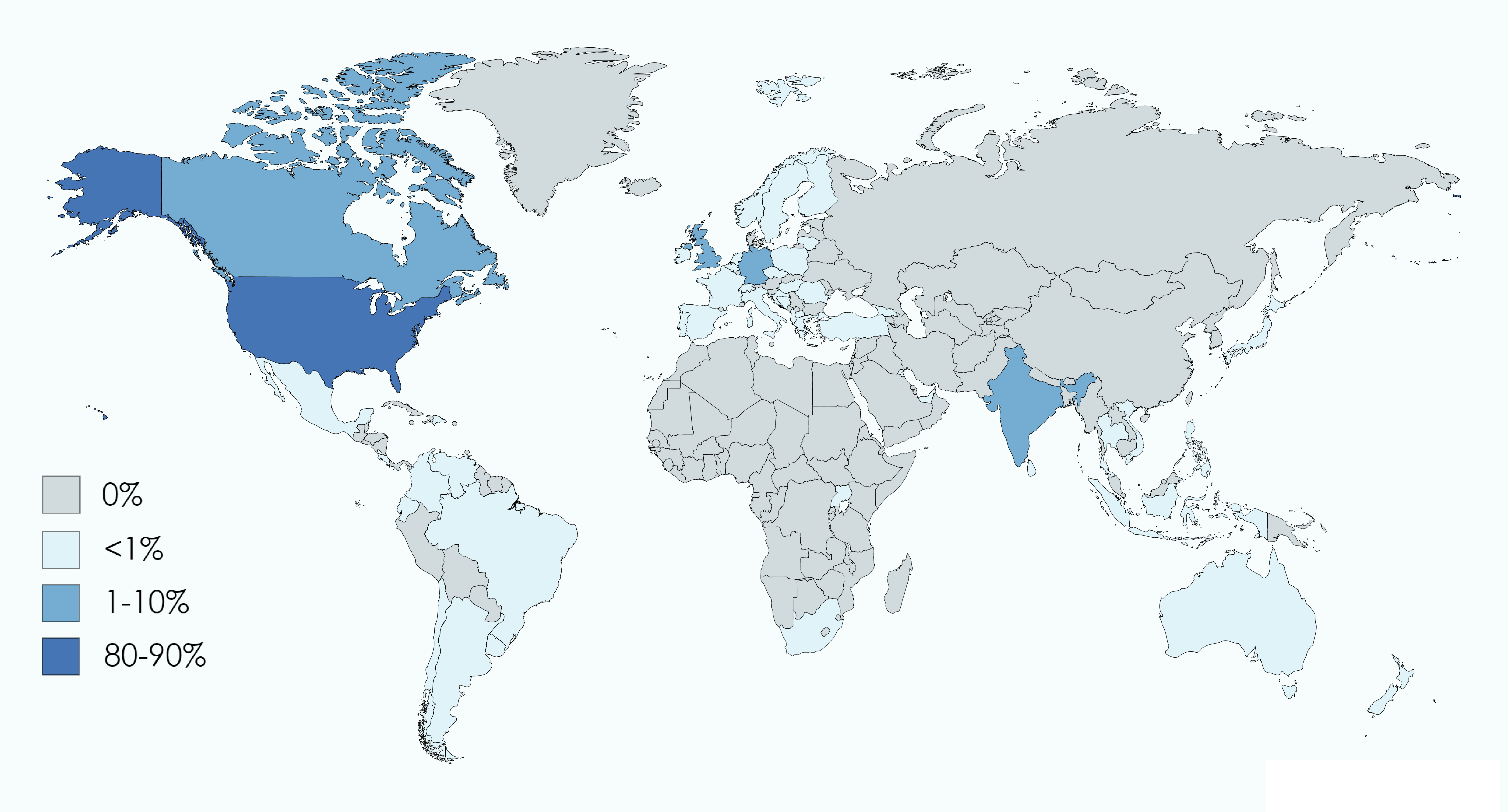}} \vspace*{-15mm}\\
\begin{minipage}[c]{0.0cm}\vspace{4.00cm}\end{minipage}&\subfloat[]{\includegraphics[width=6cm,frame]{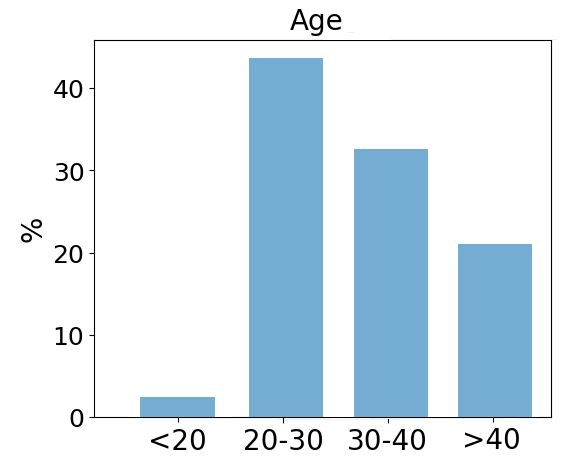}} \vspace*{-13mm}\\
\end{tabular}
\vspace*{-3mm}\caption{Participant demographics (a) Countries where participants were located; (b) Age distribution of the participants.}
\label{fig:ParticipantsDemog}
\end{figure}

The study participants were workers from AMT having approval rates exceeding 90\% on previous studies. A total of 4776 subjects took part in the experiment. The participants were from 56 different countries as highlighted in Fig. \ref{fig:ParticipantsDemog}(a) with the majority being located in the United States and India (together accounting for 91\% of the participants).  Figure \ref{fig:ParticipantsDemog}(b) shows the age distribution of the participants. About half of the participants were of each gender (46.4\% male versus 53.6\% female).

\subsubsection{Viewing Conditions}
\begin{figure*}[!htb]
\centering
\captionsetup[subfigure]{justification=centering}
\begin{tabular}{@{}c@{}c@{}c@{}c@{}c@{}c@{}}
& ~ & ~ & ~ & ~  & ~\\
\begin{minipage}[c]{0.0cm}\vspace{0.00cm}\end{minipage}&\subfloat[]{\includegraphics[height=5cm,frame]{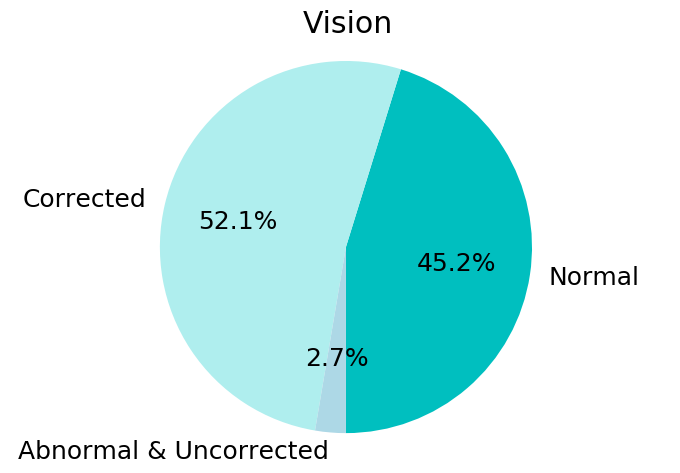}}\hspace*{1mm}
&\subfloat[]{\includegraphics[height=5cm,frame]{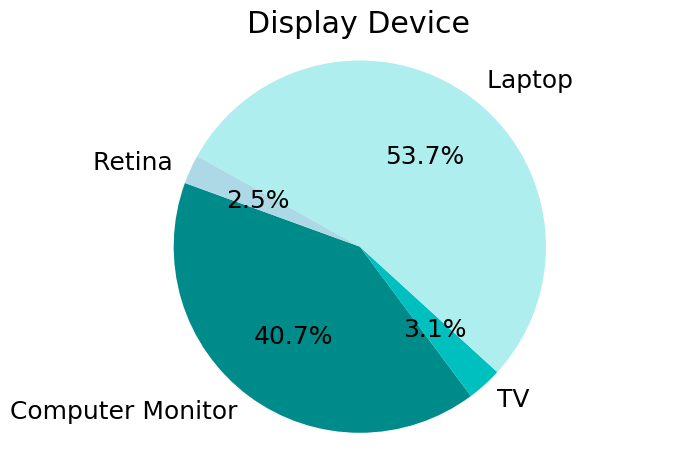}}\hspace*{1mm}
\\&\subfloat[]{\includegraphics[height=5cm,frame]{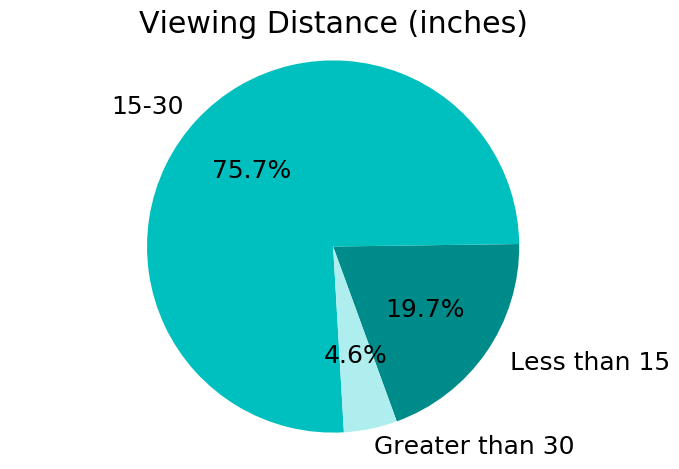}}\hspace*{1mm}
&\subfloat[]{\includegraphics[height=5cm,frame]{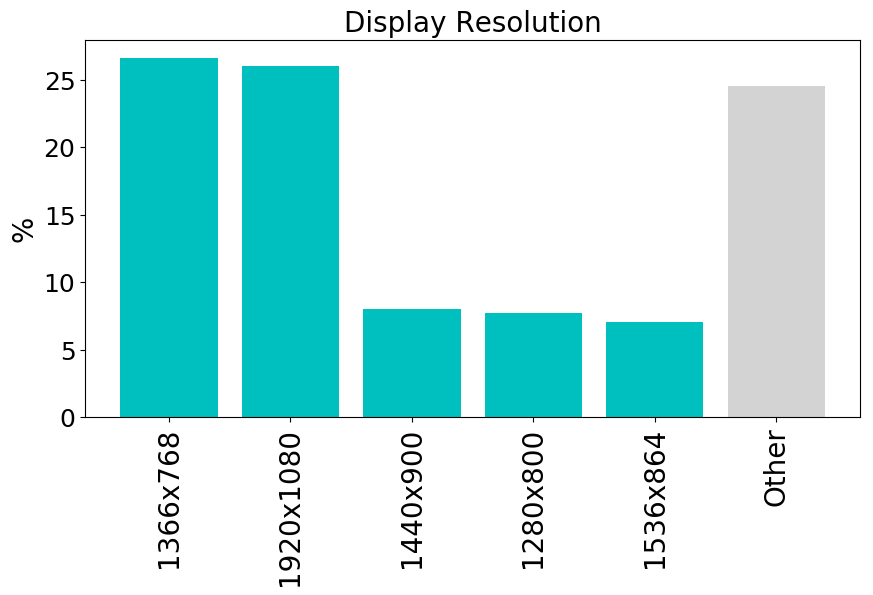}}\vspace*{-5mm}
\end{tabular}
\vspace*{4mm}\caption{Participant statistics: (a) Visual correction; (b) type of display device; (c) approximate viewing distance; (d) display resolution.}
\label{fig:ParticipantsStatistics}
\end{figure*}

As with any crowdsourced study, the participants operated under diverse viewing conditions; including locations, visual correction, viewing distances, browser, display device, resolution, ambient lighting, and so on. Figure \ref{fig:ParticipantsStatistics} presents statistics that we collected regarding some of the aspects of subject viewing. As shown in Fig. \ref{fig:ParticipantsStatistics}(a), the majority of the participants had normal or corrected-to-normal vision (e.g., glasses or contacts). A tiny percentage (2.5\%) had abnormal, uncorrected vision, and we excluded their results. The participants used mostly laptop and desktop monitors to view the videos (Fig. \ref{fig:ParticipantsStatistics}(b)), and were mostly positioned between 15 and 30 inches from the display (Fig. \ref{fig:ParticipantsStatistics}(c)).
The subjects used 83 different display resolutions which ranged between 1280$\times$720 and 3840$\times$2160 pixels, as plotted in Fig. \ref{fig:ParticipantsStatistics}(d). Of these, 31.15\% had display resolutions of at least 1920$\times$1080, while the rest had lower resolution displays.
\subsubsection{Compensation}
 On average the subjects require 16.5 minutes each to complete the study. The financial compensation given for completing the hit was one US dollar. We wanted to attract high-quality workers, hence maintaining a good AMT reputation was important. There exists a variety of forums, like TurkerNation.com, where AMT workers can share their experiences of AMT hits. These forums build a valuable sense of community among Turk workers, and helps to protect them from unfair or bad practices. We noticed that a small number of workers were uneasy about being unable to complete the study because of some of the eligibility requirements that we imposed. This was especially true when, because of hardware inadequacy, subjects were asked to return the hit during training. We notified a worker that s/he would be ineligible to continue working on the hit as soon as a problem was detected (\textit{viz.} when a video stalled $>$5 seconds or when 3 videos stalled $>$2 seconds during training). The training instructions did inform the subjects that they could be asked to return the hit if any eligibility requirement was not met. We did not compensate any ``less dedicated'' worker who skipped any video by re-enabling the controls of the video; either by using Javascript commands or by modifying browser settings, since we wanted to discourage workers from attempting such practices. Interestingly, about 2\% of the workers were ``skippers'' and were not compensated.

Adopting a strategy to reject subjects on the fly that were not providing consistent results was a more challenging issue. The previous large image quality crowdsourced study in \cite{ghadiyaram2016massive} repeated some of the images to determine how consistent a subject would rate the same distorted content, and rejected inconsistent subjects. We adopted a similar strategy, with important modifications, by repeating 4 of the videos (at random relative spacings) to measure intra-subject consistency. However this measurement problem was more complex in our study than in \cite{ghadiyaram2016massive} since hardware-related stalls; although greatly reduced in frequency, could still occur. Thus, a video and its time-displaced repeat could each present with or without stalls (of different locations and durations), thereby greatly affecting their perceived quality. We noticed that the workers were generally providing very similar ratings on videos viewed twice, when no stalls occurred, which we attribute at least in part to only including participants having high reliability scores. When stalls occurred, the results were harder to interpret. We did not want to reject the workers unfairly, hence we decided to adopt a strategy similar to that used in the crowdsourced video annotation study \cite{vondrick2013efficiently}, where the authors compensated the workers regardless of the consistency of their results, arguing that keeping a good reputation among workers was worth the overhead cost, and helped motivate the workers to produce high quality work. While we informed the workers that we reserved the right to reject poor work, we also adopted this method.
\begin{table*}[!h]
\centering
\caption{Summary of the participants that were not compensated.}
\label{notcompensated}
\begin{tabular}{|c|c|c|}
\hline
Participants Group      & Filtering                                                                                                                                                                         & Action(s)                                                                                                         \\ \hline
Ineligible Participants & \begin{tabular}[c]{@{}c@{}}Device, display, resolution and browser \\ information captured after the study overview.\\ Bandwidth and hardware tests during training.\end{tabular} & \begin{tabular}[c]{@{}c@{}}Asked to return the hit.\\ Not compensated.\\ Video scores not collected.\end{tabular} \\ \hline
Video Skippers          & Measure of the viewing duration.                                                                                                                                              & \begin{tabular}[c]{@{}c@{}}Not compensated.\\ Video scores excluded.\end{tabular}                                 \\ \hline
\end{tabular}
\end{table*}

A summary of those participants that were not compensated is given in Table 
\ref{notcompensated}.

\subsubsection{Subject Feedback}
We provided the workers with space to give comments in the exit survey. The feedback that we received was generally very positive, which suggests that the workers successfully engaged in the visual task.

Among the 4776 workers who completed the study, 32\% completed the additional comments' box. Among those, 55\% wrote that they did not have any additional comments (e.g. no comment, none, not applicable), 31\% described the test as good, nice, awesome, great, cool, enjoyable or fun. Some (13\%) of the workers provided miscellaneous comments, e.g, that they noticed that some videos repeated, or provided additional information about their display, or wondered how the results would be used, or just thanked us for the opportunity to participate.

  \section{Subjective Data Processing and Results}
Here we discuss handling of stalled videos, subject rejection, and the effects of the various experimental parameters on the study outcomes.
 \label{sec:RawScores}
\subsection{Video Stalls}
As mentioned earlier, we adopted a strategy to identify, during the training phase, those subjects that were the most susceptible to experiencing video stalls. While we were able to substantially mitigate the video stall problem, we were not able to eliminate it entirely. Although we were able to eliminate network-induced video stalls by requiring that all videos pre-loaded before display, the computational power that was available to each participants' device to play and display the videos was a stochastic resource that was a function of other processes executing in the background. While we asked the workers to close any other windows, tabs or programs, there was no way to verify whether these instructions were followed. Moreover, other necessary foreground, and background processes related to high-priority operating system tasks could affect performance. Since network connectivity can be time-varying and unpredictable, further overhead may also have weighed on processor performance during poor connections.\begin{figure}[!h]\vspace*{-0.8cm}
\centering
\captionsetup[subfigure]{labelformat=empty}
\begin{tabular}{@{}c@{}c@{}c@{}c@{}c@{}c@{}}
& ~ & ~ & ~ & ~  & ~\\
\begin{minipage}[c]{0.0cm}\vspace{-0.20cm}\end{minipage}&\subfloat[]{\includegraphics[width=7.00cm]{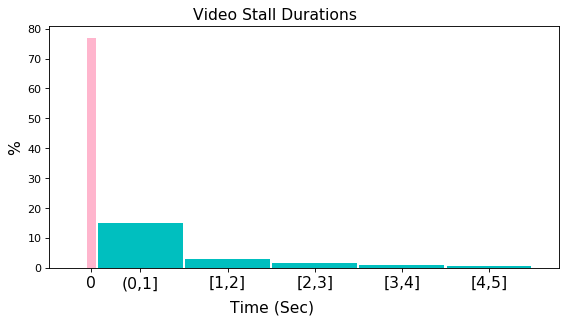}}\hspace*{0mm}\\
\end{tabular}
\vspace*{-0.5cm}
\caption{Distribution of the video stall durations.}
\label{WatchDur}
\end{figure}

Figure \ref{WatchDur} plots the distribution of the video stall durations. It can be observed 92\% of the videos had no stalls at all or had stalls that lasted for less than 1 sec. In fact,  77\% of the videos played with no stalls at all.


\subsection{Outlier Rejection}
We first rejected the ratings of those users who indicated that they wore corrective lenses, but did not wear them during the course of the experiment. This accounted for 2.5\% of all subjects. As mentioned earlier, 2\% of subjects attempted to circumvent the experiment and did not watch the videos fully; their results were also excluded.

We also excluded the ratings of users whose computation/display problems were so severe that at least 75\% of their viewed videos stalled, which eliminated ratings of 11.5\% of the subject population. The remaining subjects viewed at least 11 out of the 43 test videos (usually many more) without experiencing any stalls. For the remaining video ratings, we applied the standard BT-500-13 (Annex 2, section 2.3) \cite{rec2009bt} rejection portion on the ratings of videos, played without any stalls. We found that only 23 subjects were outliers (0.5\%) from among the entire population. This number seemed low that we also studied the intra-subject consistency. By design, each subject viewed 4 repeated videos during the test phase; we examined the differences in these pairs of scores, as follows. The average standard deviation of all non-stalled videos was about 18. We used this value as a threshold for consistency: given a non-stalled video that was viewed twice, the absolute difference in MOS of the two videos was computed. If it was below the threshold, then the rating for the video was regarded as consistent. Otherwise, it was not. We repeated this analysis across all the 4 videos across all subjects, and found that the majority ($\sim$99\%) of the subjects were self-consistent at least half of the time. It is important to emphasize that we excluded the stalled videos from the consistency analysis and when applying the subject rejection \cite{rec2009bt}, because the presence of any stalls rendered the corresponding subject ratings non-comparable.

\begin{figure}[!h]
\centering
\captionsetup[subfigure]{labelformat=empty}
\begin{tabular}{@{}c@{}c@{}c@{}c@{}c@{}c@{}}
& ~ & ~ & ~ & ~  & ~\\
\begin{minipage}[c]{0.0cm}\vspace{-0.00cm}\end{minipage}&\subfloat[]{\includegraphics[width=8.8cm]{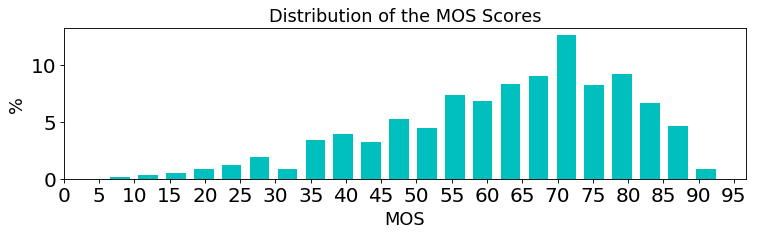}}\hspace*{-5mm}\\
\end{tabular}
\vspace{-0.50cm}
\caption{Distribution of MOS of the final set of video ratings.}
\label{fig:MOSDistributionNoStall}
\end{figure}

After rejecting the outliers, we were left with about 205 opinion scores for each video, without stalls, which is a substantial number. We computed the MOS of each video; Fig. \ref{fig:MOSDistributionNoStall} plots the histogram of these obtained MOS following all of the above-described data cleansing. It may be observed that the MOS distribution substantially spans the quality spectrum with a greater density of videos in the range 60-80. 

\subsection{Validation of Results}
\subsubsection{Golden Videos}
During the testing phase of each subject's session, 4 distorted videos taken from the LIVE VQA Database \cite{seshadrinathan2010study} - the aforementioned ``Golden Videos" - were displayed at random placements to each worker to serve as a control. The mean Spearman rank ordered correlation (SROCC) values computed between the workers' MOS on the gold standard images and the corresponding ground truth MOS values from the LIVE VQA was found to be 0.99. The mean absolute difference between the MOS values obtained from our study and the ground truth MOS values of the ``Golden Videos" was 8.5. We also conducted a paired-sampled Wilcoxon t-test, and found that the differences between these to be insignificant at $p<0.05$. A recent experiment \cite{Twitter} showed that the MOS collected in subjective studies tends to vary with the overall quality of the presented videos.  The videos in LIVE-VQC database span a wider range of quality than the LIVE VQA Database \cite{seshadrinathan2010study}, which only contains videos contaminated by only a few synthetic distortion types each at a few levels of severity. We believe that this explains the consistent shift in MOS across the 4 golden videos, when the outcomes from both experiments are compared. 

The excellent agreement between the crowdsourced scores and the laboratory MOS significantly validates our experimental protocol.

\subsubsection{Overall inter-subject consistency}
 To study overall subject consistency, we divided the opinion scores obtained on each video into two disjoint equal sets, then we computed  MOS values on each set. We conducted on all the videos, then computed the SROCC between the two sets of MOS. This experiment was repeated 100 times, and the average SROCC between the halves was found to be 0.984.

\subsection{Impact of Experimental Parameters}
\subsubsection{Number of subjects.}
\begin{figure}[!h]\vspace{-0.70cm}
 \centering
\captionsetup{justification=centering}
\begin{tabular}{@{}c@{}c@{}c@{}c@{}c@{}c@{}}
& ~ & ~ & ~\\
\begin{minipage}[c]{0.0cm}\vspace{0.00cm}\end{minipage}&\subfloat[]{\includegraphics[width=8.8cm]{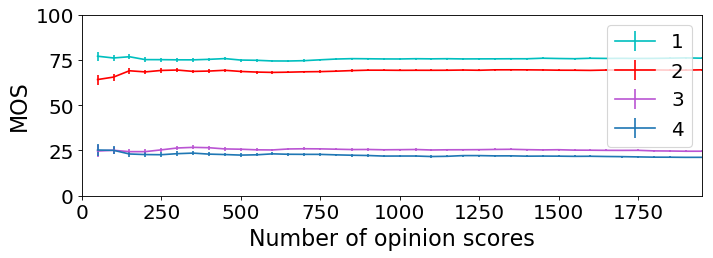}}\\ \vspace*{2mm}
\begin{minipage}[c]{0.0cm}\vspace{-0.50cm}\end{minipage}&\subfloat[]{\includegraphics[width=8.8cm]{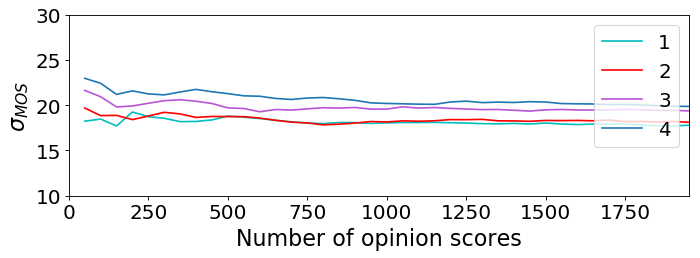}} \vspace*{-2mm}\\
\end{tabular}
\vspace*{-0.1cm}\caption{ Plots of (a) error bars of MOS; (b) standard deviation of MOS, for the set of videos viewed by all of the subjects.}
\label{fig:ParticipantsNbr}
\end{figure}

To understand the impact of the number of subjects on the obtained MOS, we considered the set of videos that were viewed by all subjects, and plotted the error bar plots of the associated MOS along with the standard deviation as a function of the number of ratings (up to 2000), as shown in Fig. \ref{fig:ParticipantsNbr}. 

We found that increasing the sample size behind beyond 200 did not improve or otherwise affect the figures. We collected slightly more than 200 opinion scores for each video (without stalls). We observed similar behaviors across the rest of the videos.

\subsubsection{Stalls.}
We computed the differential mean opinion scores (DMOS) between the non-stalled videos and the stalled videos:
\begin{equation}
DMOS=MOS_{ \;without \;  stalls}-MOS_{ \;with\;  stalls}
\label{DMOS}
\end{equation}

\begin{figure}[!h]\vspace*{-0.8cm}
\centering
\captionsetup[subfigure]{labelformat=empty}
\begin{tabular}{@{}c@{}c@{}c@{}c@{}c@{}c@{}}
& ~ & ~ & ~ & ~  & ~\\
\begin{minipage}[c]{0.0cm}\vspace{-0.20cm}\end{minipage}&\subfloat[]{\includegraphics[width=9cm]{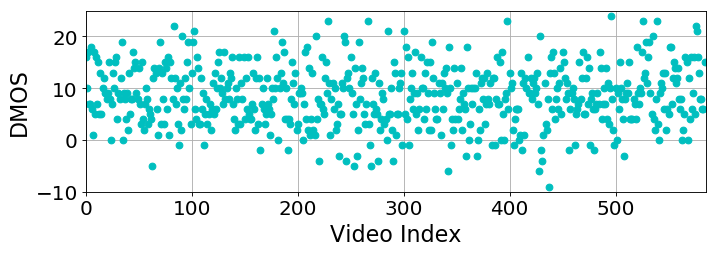}}\hspace*{0mm}\\
\end{tabular}
\vspace*{-0.5cm}
\caption{DMOS between the videos that played normally and the stalled videos.}
\label{fig:StallInfluence2}
\end{figure}

The DMOS  is plotted against the video index in Fig. \ref{fig:StallInfluence2}.

Stalls nearly always resulted in a drop of MOS (for $>$ 95\% of the videos). Since it is difficult to assert a reason for a rare small increase in MOS on a given content when stalled, we simply regard those events as noise in the data. 

While we have not included an analysis of the stalled video ratings here, we still regard the data as valuable, even though we were only able to collect the per-video total stall durations (but not the number or locations of the stalls). In future work, we plan to analyze stalled video ratings as well, with an eye towards helping guide the development of models that can account for stall occurrences when predicting the video quality.

\subsection{Worker Parameters}
\subsubsection{High vs low resolution pools}
The low resolution group (subjects with display resolutions less than 1920$\times$1080) rated 475 of the 585 videos in our database. Whereas, the high resolution pool (resolutions of at least 1920$\times$1080) rated all 585 videos. We studied the inter-subject agreement between the two groups over the common set of videos. We computed the SROCC over the MOS obtained from both groups and obtained a value of 0.97. The mean difference in MOS between the two sets was close to 1, which might be attributed to statistical noise. The high inter-subject agreement between the two groups is important, as it shows that the subjects from the high resolution pool (accounting for 31.15\% of the total population), who had seen videos of higher resolutions, did not rate the low resolution videos differently than did the low resolution pool of participants.

\subsubsection{Participants' Resolution}
As can be observed in Fig. \ref{fig:ParticipantsStatistics}(d), the two dominant resolution groups were 1366x768 and 1920$\times$1080. The other resolutions occurred less frequently (less than 10\% of the time). We studied the influence of resolution on the distribution of MOS for the two most dominant resolutions. The SROCC between the two classes was 0.95, while the mean difference in MOS between the two sets was close to zero. This result further supports our belief that the participants' display resolutions did not significantly impact the video quality ratings they supplied.

\subsubsection{Participants'  Display Devices}
Laptops and  computer monitors most often used as display devices (Fig. \ref{fig:ParticipantsStatistics}(b)). We also studied the influence of the display device, and found that it did not noticeably impact the MOS either (the SROCC between the two groups was 0.97 and the mean difference in the MOS was close to 1).

\subsubsection{Viewing Distances}
Another parameter that we studied was the reported viewing distance. There were three categories: small ($<$15 inches), medium (15-30 inches) and large ($>$30 inches) viewing distances as shown in Fig. \ref{fig:ParticipantsStatistics}(c). We found that the viewing distance had only a small effect on the distribution of MOS. The SROCC between the three categories ranged between 0.91 and 0.97, while the average difference in the MOS was less than 1.

\subsubsection{Other demographic information}
We also analyzed the impact of subjects' demographics on the distribution of MOS. First, we did not find noticeable differences between the MOS distributions across the male and female populations. The SROCC between the two gender classes was 0.97, and the average difference between the MOS was about 2; female participants tended to give slightly lower scores as compared to male participants. It is possible that this might be attributed to biological differences in the perceptual systems between of the two genders; for example, it has been reported that females are more adept at distinguishing shades of color \cite{abramov2012sex}.
\begin{table}[!h]
\centering
\caption{Spearman Correlation of the MOS distributions obtained between the different age groups.}
\label{SROCC}
\begin{tabular}{|c|c|c|c|c|}
\hline
               & \textless20 & 20-30 & 30-40 & \textgreater40 \\ \hline
\textless20    & 1           & 0.84  & 0.82  & 0.79           \\ \hline
20-30          & 0.84        & 1     & 0.97  & 0.94           \\ \hline
30-40          & 0.82        & 0.97  & 1     & 0.96           \\ \hline
\textgreater40 & 0.79        & 0.94  & 0.96  & 1              \\ \hline
\end{tabular}
\end{table}

Age did impact the MOS distribution. We compared the distributions of MOS for the four age ranges, and found that younger participants as a group delivered lower opinion scores than did older participants. These differences might be attributed to young participants having better vision \cite{rovner2002effect}, or it might related to differing expectations of younger and older viewers. As it can be observed in Table \ref{SROCC}, the larger the difference between the age groups, the lower the SROCC. The difference between the MOS distributions becomes more subtle as the difference in the age gap increases; participants younger than 20 tended to assign lower quality scores than did participants older than 40.

\section{Performance of Video Quality Predictors}
\label{Algorithms}
As mentioned earlier in the paper, we conducted this study with the aim to advance VQA research efforts, by providing a database that closely represents distorted videos encountered in the real world, along with a large number of accurate human opinions of them. In recent years, there has been numerous efforts to develop blind VQA models. Noteworthy examples include simple frame-based Natural Scene Statistics (NSS) based models, NIQE \cite{mittal2013making} and BRISQUE \cite{mittal2012no}, as well as more sophisticated predictors that incorporate more complex information such as motion. These include V-BLIINDS \cite{saad2014blind}, VIIDEO \cite{mittal2016completely}, the 3D-DCT based NR-VQA predictor described in \cite{li2016spatiotemporal}, the FC model \cite{men2017empirical}, the statistical analysis model in \cite{zhu2015no}, and the convolutional neural network model in \cite{wang2017cnn}.

To demonstrate the usefulness of our database, we evaluated the quality prediction performance of a number of leading blind VQA algorithms (whose code was publicly available). NIQE\cite{mittal2013making} and VIIDEO \cite{mittal2016completely} are training-free models capable of outputting quality scores on video. V-BLIINDS \cite{saad2014blind} and BRISQUE \cite{mittal2012no}, require training hence we learned mappings from their feature spaces to the ground truth MOS, using a support vector regressor (SVR) \cite{scholkopf2000new} that has been successfully deployed in many prior image and video quality models. We used the LIBSVM package \cite{chang2011libsvm} to implement the SVR with a radial basis function (RBF) kernel and to predict the MOS. We applied a 5-fold cross validation technique as described in  \cite{stone1974cross}. To predict quality scores over the entire database, we aggregated the predicted values obtained from each fold. The NIQE \cite{mittal2013making} features were computed on non-overlapping blocks of size 96$\times$96, then the computed NIQE distance is computed over frames and averaged over time, similar to how it was originally implemented in  V-BLIINDS \cite{saad2014blind}. BRISQUE \cite{mittal2012no} was calculated over frames and averaged in time. 

\begin{figure*}[!htb]
\centering
\captionsetup[subfigure]{justification=centering}
\begin{tabular}{@{}c@{}c@{}c@{}c@{}c@{}c@{}}
& ~ & ~ & ~ & ~  & ~\\
\begin{minipage}[c]{0.0cm}\vspace{0.00cm}\end{minipage}&\subfloat[]{\includegraphics[height=5.5cm]{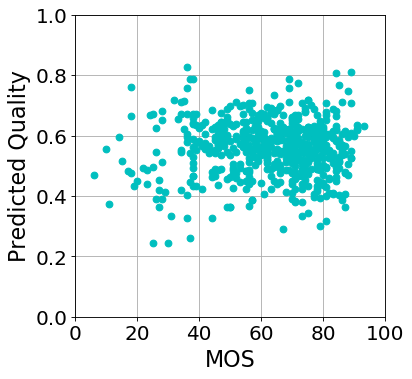}}\hspace*{1mm}
&\subfloat[]{\includegraphics[height=5.5cm]{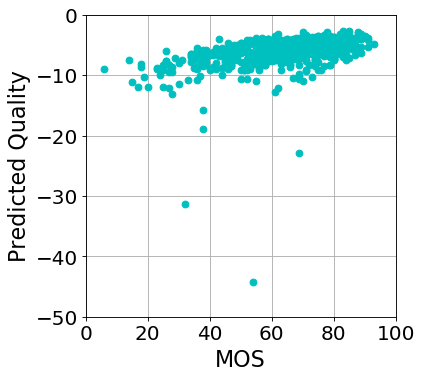}}\hspace*{1mm}
\\&\subfloat[]{\includegraphics[height=5.5cm]{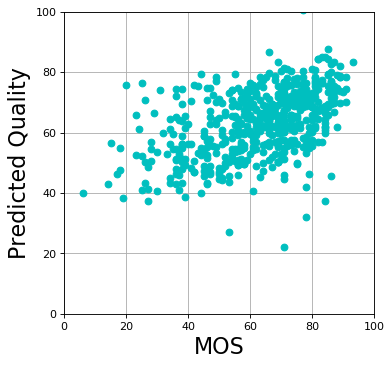}}\hspace*{1mm}
&\subfloat[]{\includegraphics[height=5.5cm]{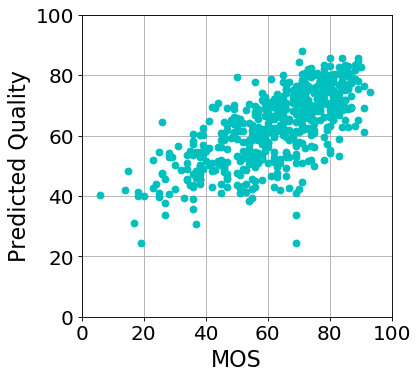}}\hspace*{1mm}
\end{tabular}
\vspace*{5mm}\caption{Scatter plots of the predicted quality scores versus MOS for four NR VQA models; (a) VIIDEO; (b) NIQE; (c) BRISQUE; (d) V-BLIINDS. }
\label{Scatters}
\end{figure*}

Figure \ref{Scatters} presents scatter plots, of NIQE \cite{mittal2013making} and VIIDEO \cite{mittal2016completely} quality predictions, and V-BLIINDS \cite{saad2014blind} and BRISQUE \cite{mittal2012no} predictions obtained after the 5-fold cross validation.
 Since NIQE provides a distance measure that increases as the video becomes more distorted, we will instead analyze the quantity -NIQE for simpler visual comparison with other models.
 As  may be observed in Fig.  \ref{Scatters}(a), the predicted VIIIDEO scores correlated poorly with the ground truth MOS, while for the other models followed regular trends against the MOS, as shown in Fig.  \ref{Scatters}(b), \ref{Scatters}(c) and \ref{Scatters}(d).

We used three performance metrics to quantify the performance of the different VQA models. First, we computed the Pearson Linear Correlation Coefficient (PLCC) between the predicted quality values and MOS distributions, after applying a non-linear mapping as prescribed in \cite{VQEG2000} to the predicted quality values. Second, we computed the Root Mean Squared Error (RMSE) between the two distributions. Finally, we computed the SROCC values between the predicted quality values and MOS values. When evaluating V-BLIINDS \cite{saad2014blind} and BRISQUE \cite{mittal2012no}, we randomly divided the videos into two disjoint sets (80\%-20\%). We used the larger set for training and the other one for testing, then we normalized our features, and fed them into the SVR module \cite{scholkopf2000new} to predict the MOS. We repeated this process 100 times, and computed the median PLCC, SROCC and RMSE values. A summary of the results obtained over all the models is given in Table \ref{resObj}. We could not run V-BLIINDS \cite{saad2014blind} successfully on all the videos, especially at lower resolutions. Unable to trace the source of this problem in the span of the current report, or resolve it, the results are reported on a subset of 553 out of the 585 videos. Note that computing the results on this subset led to a slim increase in the performance of NIQE \cite{mittal2013making} and BRISQUE \cite{mittal2012no}. VIIDEO's code \cite{mittal2016completely}'s would not run successfully on 3 videos so we report the results for it on the remaining 582 videos. The results reported for NIQE \cite{mittal2013making} and BRISQUE \cite{mittal2012no} are on the full database, since these algorithms could run successfully on all the videos. As may be observed, V-BLIINDS supplied the best performance in terms of the three performance metrics. However, there remain ample room for improvement, suggesting the need for developing better NR VQA models, capable of better assessing authentic, real-world video distortions.

\begin{table}[!htb]
\centering
\caption{Performance Metrics Measured on the Compared VQA Models.}
\label{resObj}
\begin{tabular}{|c|c|c|c|}
\hline
        & PLCC & SROCC & RMSE \\ \hline
VIIDEO  \cite{mittal2016completely}& 0.1366 & -0.0293     & 16.851   \\ \hline
NIQE \cite{mittal2013making}    & 0.5832 & 0.5635  & 13.857   \\ \hline
BRISQUE \cite{mittal2012no} & 0.6456 & 0.6072  & 12.908   \\ \hline
V-BLIINDS \cite{saad2014blind}& 0.7196 & 0.7083  & 11.478   \\ \hline
\end{tabular}
\end{table}

\section{Concluding Remarks and Future Work}
\label{Conclusion}
We have described the construction of a new ``in the wild'' video quality database, LIVE-VQC, containing 585 videos of unique contents, and impaired by authentic distortion combinations, captured by 80 users around the globe using 43 different device models. We also designed and built a crowdsourced framework to collect more than 205000 online opinion scores of the quality of these videos. The AMT subjects who participated were located in 56 different countries, represented genders about equally, and spanned a wide range of ages. The significant diversity of the subject pool raised many technical challenges owing to widely differing viewing conditions and resources. However, the framework we built proved to be robust against the many variables affecting the video rating process. 
While the VQA models that we tested did not perform particularly well on the new database, this was not unexpected as existing NR VQA models were not been adequately engineered to deal with so many real-world distortions. Moving forward, our work shows that improved blind VQA models better capable of predicting the quality of real-world videos are needed.

\section*{Acknowledgments}
The authors would like to thank the numerous video data contributors for their extensive efforts in providing us the video content that was used to build our database. This work would not have been possible without their help.

\bibliographystyle{IEEEtran}
\bibliography{refs}

\begin{thebibliography}{10}
\providecommand{\url}[1]{#1}
\csname url@samestyle\endcsname
\providecommand{\newblock}{\relax}
\providecommand{\bibinfo}[2]{#2}
\providecommand{\BIBentrySTDinterwordspacing}{\spaceskip=0pt\relax}
\providecommand{\BIBentryALTinterwordstretchfactor}{4}
\providecommand{\BIBentryALTinterwordspacing}{\spaceskip=\fontdimen2\font plus
\BIBentryALTinterwordstretchfactor\fontdimen3\font minus
  \fontdimen4\font\relax}
\providecommand{\BIBforeignlanguage}[2]{{%
\expandafter\ifx\csname l@#1\endcsname\relax
\typeout{** WARNING: IEEEtran.bst: No hyphenation pattern has been}%
\typeout{** loaded for the language `#1'. Using the pattern for}%
\typeout{** the default language instead.}%
\else
\language=\csname l@#1\endcsname
\fi
#2}}
\providecommand{\BIBdecl}{\relax}
\BIBdecl

\bibitem{bergman_2017}
\BIBentryALTinterwordspacing
S.~Bergman, ``We spend a billion hours a day on {Y}ou{T}ube, more than
  {N}etflix and {F}acebook video combined,'' \emph{Forbes}, Feb 2017, accessed
  on Jan. 2 2018. [Online]. Available:
  \url{https://www.forbes.com/sites/sirenabergman/2017/02/28/we-spend-a-billion-hours-a-day-on-youtube-more-than-netflix-and-facebook-video-combined/\#38c001ba5ebd}
\BIBentrySTDinterwordspacing

\bibitem{goodrow_2017}
\BIBentryALTinterwordspacing
C.~Goodrow, ``You know what’s cool? {A} billion hours,'' Feb 2017, accessed
  on Jan. 2 2018. [Online]. Available:
  \url{https://youtube.googleblog.com/2017/02/you-know-whats-cool-billion-hours.html}
\BIBentrySTDinterwordspacing

\bibitem{netflix}
\BIBentryALTinterwordspacing
C.~Riley, ``Brits can't get enough of {N}etflix and {A}mazon {P}rime,'' Jul
  2018, accessed on Jul. 25 2018. [Online]. Available:
  \url{https://money.cnn.com/2018/07/18/technology/amazon-netflix-uk-subscribers/index.html}
\BIBentrySTDinterwordspacing

\bibitem{cisco_2017}
\BIBentryALTinterwordspacing
``Cisco visual networking index: Forecast and methodology, 2016–2021,''
  \emph{Cisco}, Sep 2017. [Online]. Available:
  \url{https://www.cisco.com/c/en/us/solutions/collateral/service-provider/visual-networking-index-vni/complete-white-paper-c11-481360.html}
\BIBentrySTDinterwordspacing

\bibitem{seshadrinathan2010study}
K.~Seshadrinathan, R.~Soundararajan, A.~C. Bovik, and L.~K. Cormack, ``Study of
  subjective and objective quality assessment of video,'' \emph{IEEE Trans.
  Image Process.}, vol.~19, no.~6, pp. 1427--1441, 2010.

\bibitem{de2010h}
F.~De~Simone, M.~Tagliasacchi, M.~Naccari, S.~Tubaro, and T.~Ebrahimi, ``A {H}.
  264/{A}{V}{C} video database for the evaluation of quality metrics,''
  \emph{Intern. Conf. Acous. Sp. Sign. Process. (ICASSP)}, pp. 2430--2433,
  2010.

\bibitem{chen2014modeling}
C.~Chen, L.~K. Choi, G.~de~Veciana, C.~Caramanis, R.~W. Heath, and A.~Bovik,
  ``Modeling the time varying subjective quality of {H}{T}{T}{P} video streams
  with rate adaptations,'' \emph{IEEE Trans. on Image Process.}, vol.~23,
  no.~5, pp. 2206--2221, 2014.

\bibitem{moorthy2012video}
A.~K. Moorthy, L.~K. Choi, A.~C. Bovik, and G.~De~Veciana, ``Video quality
  assessment on mobile devices: Subjective, behavioral and objective studies,''
  \emph{IEEE J. Select. Topics Sign. Process.}, vol.~6, no.~6, pp. 652--671,
  2012.

\bibitem{keimel2010visual}
C.~Keimel, J.~Habigt, T.~Habigt, M.~Rothbucher, and K.~Diepold, ``Visual
  quality of current coding technologies at high definition {I}{P}{T}{V}
  bitrates,'' \emph{IEEE Int'l Wkshp. Multidim. Sign. Process.}, pp. 390--393,
  2010.

\bibitem{keimel2012tum}
C.~Keimel, A.~Redl, and K.~Diepold, ``The {T}{U}{M} high definition video
  datasets,'' \emph{Int'l Wkshp. Qual. Multim. Exper.}, pp. 97--102, 2012.

\bibitem{lin2015mcl}
J.~Y. Lin, R.~Song, C.-H. Wu, T.~Liu, H.~Wang, and C.-C.~J. Kuo, ``M{C}{L}-{V}:
  A streaming video quality assessment database,'' \emph{J. Vis. Commun. Image
  Repres.}, vol.~30, pp. 1--9, 2015.

\bibitem{zhang_2016}
H.~Wang, I.~Katsavounidis, J.~Zhou, J.~Park, S.~Lei, X.~Zhou, M.-O. Pun,
  X.~Jin, R.~Wang, X.~Wang \emph{et~al.}, ``Videoset: A large-scale compressed
  video quality dataset based on jnd measurement,'' \emph{J. of Vis. Comm. Im.
  Rep.}, vol.~46, pp. 292--302, 2017.

\bibitem{ghadiyaram2017capture}
D.~Ghadiyaram, J.~Pan, A.~C. Bovik, A.~K. Moorthy, P.~Panda, and K.-C. Yang,
  ``In-capture mobile video distortions: A study of subjective behavior and
  objective algorithms,'' \emph{IEEE Trans. Circ. Syst. Video Technol.}, 2018.

\bibitem{nuutinen2016cvd2014}
M.~Nuutinen, T.~Virtanen, M.~Vaahteranoksa, T.~Vuori, P.~Oittinen, and
  J.~H{\"a}kkinen, ``C{V}{D}2014 a database for evaluating no-reference video
  quality assessment algorithms,'' \emph{IEEE Trans. Image Process.}, vol.~25,
  no.~7, pp. 3073--3086, 2016.

\bibitem{russakovsky2015imagenet}
O.~Russakovsky, J.~Deng, H.~Su, J.~Krause, S.~Satheesh, S.~Ma, Z.~Huang,
  A.~Karpathy, A.~Khosla, M.~Bernstein \emph{et~al.}, ``Imagenet large scale
  visual recognition challenge,'' \emph{Intern. J. Comp. Vis.}, vol. 115,
  no.~3, pp. 211--252, 2015.

\bibitem{redi2013crowdsourcing}
J.~A. Redi, T.~Ho{\ss}feld, P.~Korshunov, F.~Mazza, I.~Povoa, and C.~Keimel,
  ``Crowdsourcing-based multimedia subjective evaluations: a case study on
  image recognizability and aesthetic appeal,'' \emph{ACM Intern. Wkshp Crowds.
  Multim.}, pp. 29--34, 2013.

\bibitem{ghadiyaram2016massive}
D.~Ghadiyaram and A.~C. Bovik, ``Massive online crowdsourced study of
  subjective and objective picture quality,'' \emph{IEEE Trans. Image
  Process.}, vol.~25, no.~1, pp. 372--387, 2016.

\bibitem{chen2010quadrant}
K.~T. Chen, C.~J. Chang, C.~C. Wu, Y.~C. Chang, and C.~L. Lei, ``Quadrant of
  euphoria: {A} crowdsourcing platform for {Q}o{E} assessment,'' \emph{IEEE
  Net.}, vol.~24, no.~2, 2010.

\bibitem{seufert2016one}
M.~Seufert and T.~Ho{\ss}feld, ``One shot crowdtesting: Approaching the
  extremes of crowdsourced subjective quality testing,'' \emph{Wkshp Perc.
  Qual. Sys. (PQS 2016)}, pp. 122--126, 2016.

\bibitem{keimel2012qualitycrowd}
C.~Keimel, J.~Habigt, C.~Horch, and K.~Diepold, ``Qualitycrowd a framework for
  crowd-based quality evaluation,'' \emph{Pict. Cod. Symp. (PCS), 2012}, pp.
  245--248, 2012.

\bibitem{hosu2017konstanz}
V.~Hosu, F.~Hahn, M.~Jenadeleh, H.~Lin, H.~Men, T.~Szir{\'a}nyi, S.~Li, and
  D.~Saupe, ``The konstanz natural video database (konvid-1k),'' \emph{Qual.
  Mult. Exp. (QoMEX)}, pp. 1--6, 2017.

\bibitem{rec2009bt}
``Methodology for the subjective assessment of the quality of television
  pictures.''\hskip 1em plus 0.5em minus 0.4em\relax ITU-R Rec. BT. 500-13,
  2012.

\bibitem{hossfeld2014best}
T.~Hossfeld, C.~Keimel, M.~Hirth, B.~Gardlo, J.~Habigt, K.~Diepold, and
  P.~Tran-Gia, ``Best practices for {Q}o{E} crowdtesting: {Q}o{E} assessment
  with crowdsourcing,'' \emph{IEEE Trans. Multim.}, vol.~16, no.~2, pp.
  541--558, 2014.

\bibitem{figuerola2013assessing}
{\'O}.~Figuerola~Salas, V.~Adzic, A.~Shah, and H.~Kalva, ``Assessing internet
  video quality using crowdsourcing,'' \emph{Proc. ACM Int'; Wkshp. Crowd.
  Multim.}, pp. 23--28, 2013.

\bibitem{shahid2014crowdsourcing}
M.~Shahid, J.~S{\o}gaard, J.~Pokhrel, K.~Brunnstr{\"o}m, K.~Wang, S.~Tavakoli,
  and N.~Gracia, ``Crowdsourcing based subjective quality assessment of
  adaptive video streaming,'' \emph{Wkshp. Qual. Multim. Exper.}, pp. 53--54,
  2014.

\bibitem{chen2015qos}
Y.~Chen, K.~Wu, and Q.~Zhang, ``From {Q}o{S} to {Q}o{E}: A tutorial on video
  quality assessment,'' \emph{IEEE Comm. Surv. Tutorials}, vol.~17, no.~2, pp.
  1126--1165, 2015.

\bibitem{rainer2014quality}
B.~Rainer and C.~Timmerer, ``Quality of experience of web-based adaptive
  {H}{T}{T}{P} streaming clients in real-world environments using
  crowdsourcing,'' \emph{ACM Wkshp. Desig. Quality Deployment Adapt. Video
  Streaming}, pp. 19--24, 2014.

\bibitem{linkVQC}
\BIBentryALTinterwordspacing
Z.~Sinno and A.~C. Bovik, ``Live video quality challenge ({V}{Q}{C})
  database,'' \emph{Laboratory for Image and Video Engineering}, Aug 2018,
  accessed on Aug. 11 2018. [Online]. Available:
  \url{http://live.ece.utexas.edu/research/LIVEVQC/index.html}
\BIBentrySTDinterwordspacing

\bibitem{Gartner}
\BIBentryALTinterwordspacing
L.~Gausduff and A.~A. Forni, ``Gartner says worldwide sales of smartphones grew
  7 percent in the fourth quarter of 2016,'' Feb 2017, accessed on Jan. 2 2018.
  [Online]. Available: \url{https://www.gartner.com/newsroom/id/3609817l}
\BIBentrySTDinterwordspacing

\bibitem{Resolutions}
\BIBentryALTinterwordspacing
``Browser market share worldwide {N}ovember 2016$-${D}ecember 2017,'' Dec 2017,
  accessed on Jan. 2 2018. [Online]. Available:
  \url{http://gs.statcounter.com/}
\BIBentrySTDinterwordspacing

\bibitem{sinno2018icip}
Z.~Sinno and A.~C. Bovik, ``Large scale subjective video quality study,'' in
  \emph{IEEE Intern. Conf. Image Process. (ICIP)}, Oct. 2018, accepted.

\bibitem{vondrick2013efficiently}
C.~Vondrick, D.~Patterson, and D.~Ramanan, ``Efficiently scaling up
  crowdsourced video annotation,'' \emph{Int. J. Comp. Vis.}, vol. 101, no.~1,
  pp. 184--204, 2013.

\bibitem{Twitter}
\BIBentryALTinterwordspacing
S.~Van~Leuven, ``Measuring perceived video quality on mobile devices,''
  \emph{Twitter's Engineering Blog}, Jul 2018, accessed on Aug. 9 2018.
  [Online]. Available:
  \url{https://blog.twitter.com/engineering/en_us/topics/insights/2018/videoqualityonmobile.html}
\BIBentrySTDinterwordspacing

\bibitem{abramov2012sex}
I.~Abramov, J.~Gordon, O.~Feldman, and A.~Chavarga, ``Sex and vision {I}{I}:
  color appearance of monochromatic lights,'' \emph{Biol. Sex Diff.}, vol.~3,
  no.~1, p.~21, 2012.

\bibitem{rovner2002effect}
B.~W. Rovner, R.~J. Casten, and W.~S. Tasman, ``Effect of depression on vision
  function in age-related macular degeneration,'' \emph{Arch. Opht.}, vol. 120,
  no.~8, pp. 1041--1044, 2002.

\bibitem{mittal2013making}
A.~Mittal, R.~Soundararajan, and A.~C. Bovik, ``Making a “completely blind”
  image quality analyzer,'' \emph{IEEE Sign. Process, Lett.}, vol.~20, no.~3,
  pp. 209--212, 2013.

\bibitem{mittal2012no}
A.~Mittal, A.~K. Moorthy, and A.~C. Bovik, ``No-reference image quality
  assessment in the spatial domain,'' \emph{IEEE Trans. on Image Proc.},
  vol.~21, no.~12, pp. 4695--4708, 2012.

\bibitem{saad2014blind}
M.~A. Saad, A.~C. Bovik, and C.~Charrier, ``Blind prediction of natural video
  quality,'' \emph{IEEE Trans. Image Process.}, vol.~23, no.~3, pp. 1352--1365,
  2014.

\bibitem{mittal2016completely}
A.~Mittal, M.~A. Saad, and A.~C. Bovik, ``A completely blind video integrity
  oracle,'' \emph{IEEE Trans. Image Process.}, vol.~25, no.~1, pp. 289--300,
  2016.

\bibitem{li2016spatiotemporal}
X.~Li, Q.~Guo, and X.~Lu, ``Spatiotemporal statistics for video quality
  assessment,'' \emph{IEEE Trans. Image Process.}, vol.~25, no.~7, pp.
  3329--3342, 2016.

\bibitem{men2017empirical}
H.~Men, H.~Lin, and D.~Saupe, ``Empirical evaluation of no-reference vqa
  methods on a natural video quality database,'' \emph{Wkshp. Qual. Multim.
  Exper.}, pp. 1--3, 2017.

\bibitem{zhu2015no}
K.~Zhu, C.~Li, V.~Asari, and D.~Saupe, ``No-reference video quality assessment
  based on artifact measurement and statistical analysis,'' \emph{IEEE Trans.
  on Circ. Sys. Vid. Tech.}, vol.~25, no.~4, pp. 533--546, 2015.

\bibitem{wang2017cnn}
C.~Wang, L.~Su, and Q.~Huang, ``Cnn-mr for no reference video quality
  assessment,'' \emph{Inform. Sci. Contr. Eng. (ICISCE)}, pp. 224--228, 2017.

\bibitem{scholkopf2000new}
B.~Sch{\"o}lkopf, A.~J. Smola, R.~C. Williamson, and P.~L. Bartlett, ``New
  support vector algorithms,'' \emph{Neur. Comp.}, vol.~12, no.~5, pp.
  1207--1245, 2000.

\bibitem{chang2011libsvm}
C.~C. Chang and C.~J. Lin, ``{LIBSVM}: A library for support vector machines,''
  \emph{ACM Trans. Intell. Sys. Tech.}, vol.~2, no.~3, p.~27, 2011.

\bibitem{stone1974cross}
M.~Stone, ``Cross-validatory choice and assessment of statistical
  predictions,'' \emph{J. Royal Stat. Soc. Ser. B (Methodol.)}, pp. 111--147,
  1974.

\bibitem{VQEG2000}
``Final report from the video quality experts group on the validation of
  objective models of video quality assessment.''\hskip 1em plus 0.5em minus
  0.4em\relax Video Quality Expert Group (VQEG), 2000.

\end{thebibliography}

~~~
\begin{wrapfigure}{l}{39mm} 
    \includegraphics[width=1.6in,clip,keepaspectratio]{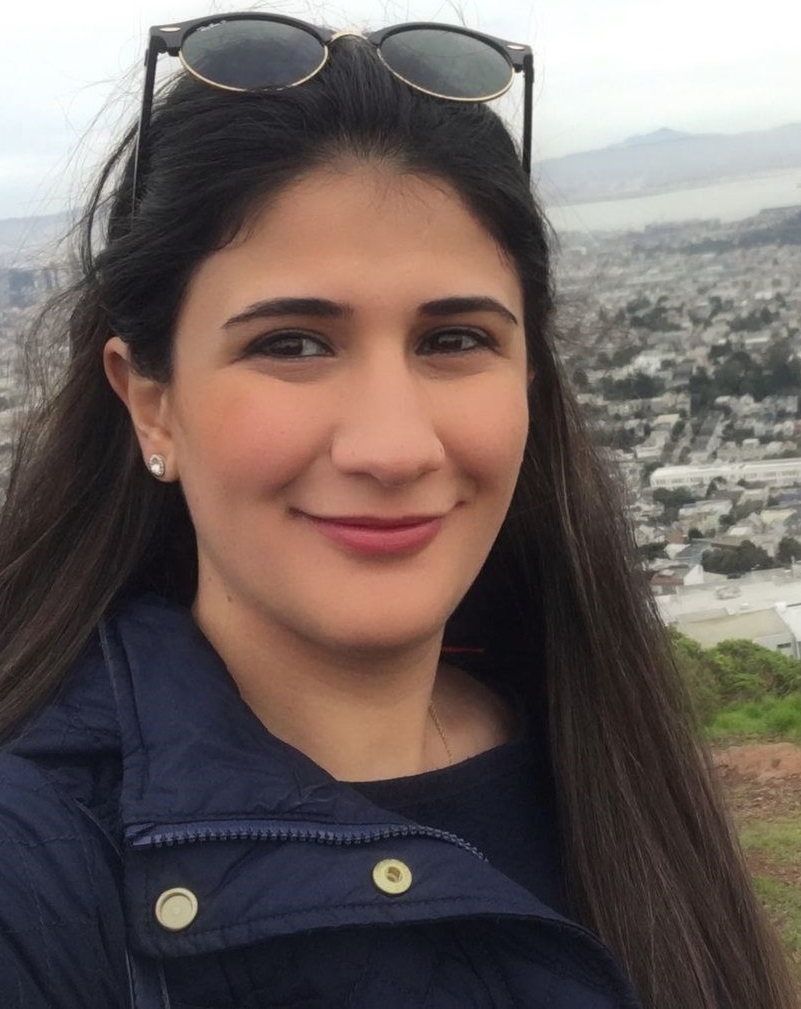}
  \end{wrapfigure}\par
  \textbf{Zeina Sinno}
received the B.E. degree (Hons.) in
electrical and computer engineering with a minor in
mathematics from the American University of Beirut
in 2013 and the M.S. in electrical and computer
engineering from The University of Texas at Austin
in 2015, where she is currently pursuing the Ph.D.
degree with the Laboratory for Image and Video
Engineering. Her research interests focus on image
and video processing, and machine learning.\par
~~\par
~~~~~~~~~

\begin{wrapfigure}{l}{39mm} 
    \includegraphics[width=1.6in,clip,keepaspectratio]{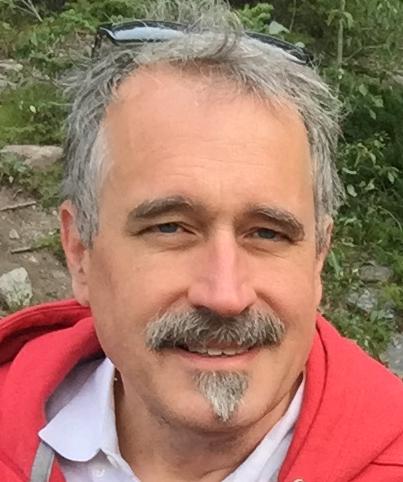}
  \end{wrapfigure}\par
  \textbf{Alan Conrad Bovik}
(F’95) received the B.S., M.S.,
and Ph.D. degrees in electrical and computer engineering
from the University of Illinois in 1980, 1982,
and 1984, respectively. He is currently the Cockrell
Family Regents Endowed Chair Professor with The
University of Texas at Austin. His books include the
Handbook of Image and Video Processing, Modern
Image Quality Assessment, and The Essential Guides
to Image and Video Processing. His research interests
are digital video, image processing, and visual
perception. He was a recipient of the 2015 Primetime
Emmy Award for outstanding achievement in engineering development from
the Television Academy, the 2017 Edwin H. Land Medal from the Optical
Society of America, the 2019 IEEE Fourier Award, and the Society Award
from the IEEE Signal Processing Society. He has also received about 10 journal
Best Paper Awards including the 2016 IEEE Signal Processing Society
Sustained Impact Award. He also created/Chaired the IEEE International
Conference on Image Processing, which was first held in Austin, TX, USA,
1994. He Co-Founded and was the longest-serving Editor-in-Chief of the
IEEE Transactions on Image Processing.\par

\end{document}